\documentstyle[12pt,aaspp4,flushrt]{article}

\begin{document}

\title
  {The Power Spectrum of Galaxy Clustering in the Las Campanas Redshift Survey}
\author{Huan Lin \altaffilmark{1} and Robert P. Kirshner}
\affil{Harvard-Smithsonian Center for Astrophysics, 60 Garden St.,
       Cambridge, MA 02138, USA \\
       hlin@cfa.harvard.edu, kirshner@cfa.harvard.edu}
\author{Stephen A. Shectman and Stephen D. Landy}
\affil{Carnegie Observatories, 813 Santa Barbara St., Pasadena, CA
       91101, USA \\
       shec@ociw.edu, landy@ociw.edu}
\author{Augustus Oemler}
\affil{Dept. of Astronomy, Yale University, 
       New Haven, CT 06520-8101, USA \\
       oemler@astro.yale.edu}
\author{Douglas L. Tucker}
\affil{Astrophysikalisches Institut Potsdam, An der Sternwarte 16,
       D-14482 Potsdam, Germany \\
       dtucker@aip.de}
\and
\author{Paul L. Schechter}
\affil{Dept. of Physics, Massachusetts Institute of Technology,
       Cambridge, MA 02139, USA \\
       schech@achernar.mit.edu}

\altaffiltext{1}{Present Affiliation: Dept. of Astronomy, University of
    Toronto, 60 St. George St., Toronto, ON M5S 3H8, Canada, 
    lin@astro.utoronto.ca}

\clearpage

\begin{abstract}

The Las Campanas Redshift Survey (LCRS) contains 23697
galaxies, with an average redshift $z = 0.1$, distributed over six
$1.5\arcdeg$ by $80\arcdeg$ slices in the North and South galactic caps.
We have computed the power spectrum $P(k)$ for magnitude-limited samples of 
LCRS galaxies over wavelengths $\lambda = 2\pi / k = 5 - 400\ h^{-1}$~Mpc. 
The LCRS $P(k)$ may be approximated as
$\propto k^{-1.8 \pm 0.1}$ for small scales $\lambda = 5 - 30\ h^{-1}$~Mpc, 
changing to
$\propto k^{1 \pm 1}$ for large scales $\lambda \approx 200 - 400\ h^{-1}$~Mpc.
The overall amplitude corresponds to $\sigma_8 = 1.0 \pm 0.1$ 
in redshift space.

Comparisons to the power spectra of other redshift surveys will be presented;
the LCRS results agree best with those from the combined 
Center for Astrophysics (CfA2) and Southern Sky redshift surveys (SSRS2).
For $\lambda \gtrsim \ 100\ h^{-1}$~Mpc, the LCRS results are consistent with
those of other surveys, given the large errors among all the surveys on
these scales. 
For $\lambda \lesssim \ 100\ h^{-1}$~Mpc, the LCRS $P(k)$ is well determined
and similar in shape to the $P(k)$ of other surveys, but with an 
amplitude differing from
some of the other samples, possibly because of inherent clustering differences
among different types of galaxies. In particular, 
power spectrum measurements for
volume-limited LCRS samples show that galaxies brighter than about
$M^* - 1$ 
appear about 50\% more strongly clustered than those fainter.
Also, a sample of LCRS emission
galaxies shows 30\% weaker clustering than the full LCRS sample. 

Comparisons to N-body
models show that the LCRS power spectrum lies intermediate between that of 
a standard flat $\Omega_0 h = 0.5$ cold dark matter (CDM)
model and an open $\Omega_0 h = 0.2$ model, both normalized to $\sigma_8 =
1$ for galaxies. On large scales 
$\lambda \gtrsim 40\ h^{-1}$~Mpc, we have fit the LCRS results to
various linear CDM models, and find that a number of them could meet
the constraints set by the LCRS power
spectrum, the Hubble constant range $0.5 \lesssim h \lesssim 0.8$, the 
abundance of galaxy clusters, and the reasonable assumption that LCRS galaxies
are roughly unbiased tracers of the mass, relative to the
normalization provided by the 4-year COBE DMR data. The possibilites include
open CDM or flat non-zero cosmological-constant CDM models with $\Omega_0
\approx 0.4-0.6$ and shape parameter $\Gamma \approx \Omega_0 h 
\approx 0.2-0.3$, as well
as flat $\Omega_0 = 1$ models with massive neutrino density
$\Omega_\nu \approx 0.2-0.3$ or a spectral tilt $n \approx 0.7-0.8$.
 
\end{abstract}

\keywords{cosmology: observations --- galaxies: clustering ---
large-scale structure of universe}

\clearpage

\section{Introduction} \label{intro}

The power spectrum of density fluctuations is an important fundamental 
quantity of interest for the problem of structure formation in cosmology.
For example, for Gaussian density fluctuations, as may arise out
of inflationary scenarios for the origin of the universe (\cite{bar83}),
the power spectrum provides a complete description of the initial fluctuations.
The subsequent development and evolution of structure via gravitational
instability, in the context of a dark matter dominated universe and models
of biased galaxy formation (\cite{bar86}), have received much theoretical 
attention in the past decade, both analytically and through the use of
large numerical N-body simulations (e.g., \cite{dav85}; \cite{par91};
\cite{zurek94}). On the observational front, the detection of microwave 
background anisotropies by the COBE DMR experiment (\cite{smo92}) provided
important constraints on the primordial power spectrum on horizon
scales. The clustering of galaxies on smaller scales and the interpretation
of the observations to reveal the matter content of the universe, to 
elucidate the process of structure formation, and to connect to the primordial
fluctuations seen by COBE, remain important problems in observational 
cosmology (see e.g. \cite{lyt94} for a review).

Much recent attention has been paid toward determination of the power spectrum
of galaxy clustering in a variety of galaxy surveys, including the 
CfA (\cite{vog92}; \cite{par94}), SSRS (\cite{par92}; \cite{daC94}),
IRAS 1.2 Jy (\cite{fis93}), IRAS QDOT (\cite{fel94}), and APM 
(\cite{bau93}, 1994) surveys.
In brief, the power spectra of these surveys have
appeared inconsistent with predictions of the ``standard'' biased cold dark 
matter (CDM) model of structure formation with 
$\Omega_0 h = 0.5$ (\cite{blu84}), while an unbiased $\Omega_0 h \approx 0.2$ model
with more large scale power agrees better with the observations
(e.g. \cite{daC94}). (We express the Hubble constant as 
$H_0 = 100 \ h$~km s$^{-1}$ Mpc$^{-1}$, and will use $h = 1$ unless
otherwise indicated.)
In this paper we present the power spectrum for galaxy samples drawn from
the Las Campanas Redshift Survey (LCRS), an optically-selected 
survey of 23697 galaxies with an average redshift $z = 0.1$. 
The large sample size and extent of our survey allow us to examine the
power spectrum up to wavelengths of $\approx 400\ h^{-1}$ Mpc, and to 
provide measurements independent of previous results 
for the purpose of comparing against cosmological models.
In particular, measurements of the power spectrum on the largest scales 
$\lambda \gtrsim 100 \ h^{-1}$~Mpc are especially interesting, as
we expect the power spectrum to peak there and begin its turnover toward the
primordial spectrum constrained by COBE and other microwave background observations.
The precise amplitude and shape of the power spectrum on 
large scales will provide important clues in discriminating among
cosmological models.

A detailed description of the Las Campanas survey is given in 
Shectman et al.\ (1996),
and additional particulars may be found in Shectman et al.\ 
(1992, 1995), Tucker (1994), Lin et al.\ (1996) and Oemler et al.\ (1993). Here we briefly
describe the main survey parameters. The survey geometry is that of six
$1.5\arcdeg \times 80\arcdeg$ ``slices'' (declination by right ascension), 
three each in the North and South galactic caps. Figure~\ref{figslices}
shows the LCRS galaxy distribution and clearly illustrates the
striking pattern of clusters, filaments, walls and voids that is
present. The first 20\% of the data was obtained
using a 50-object fiber-optic spectrograph, and the remaining 80\% of
the data was taken with a 112-object system. 
The nominal isophotal magnitude
limits for the 50-fiber data were $16.0 \leq m < 17.3$ (``hybrid''
Kron-Cousins $R$ magnitudes), 
and an additional cut was applied which
excluded the lowest 20\% of galaxies by central surface brightness. 
For the 112-fiber data, the nominal magnitude limits were $15.0 \leq m < 17.7$, with
exclusion of just the lowest 4-9\% of galaxies by surface brightness.
The survey photometric limits were chosen so that there would be typically more targets 
per field than available fibers, and we selected targets at random among
those that met the selection criteria. 
The survey slices were built up by observing $1.5\arcdeg \times 1.5\arcdeg$ 
fields, one at a time, with a maximum of 50 or 112 galaxies observed per field.
Because we generally do not re-observe any of 
our fields, we must keep track of the variable field-to-field sampling 
fractions $f$ in our subsequent statistical analyses. The average
sampling fraction is 70\% for the 112-fiber data and 58\% for the 50-fiber data.
Also, mechanical constraints prevent two object fibers in a single
spectroscopic field from approaching closer than $55\arcsec$,
introducing an additional geometric selection effect. We will find
below that the various sampling, photometric, and geometric selection
effects in our survey do not significantly affect the power spectrum results.

In \S~\ref{ps:methods} we detail our power spectrum estimation techniques and
verify them on N-body simulations. In \S~\ref{pslcg} we present the power
spectra of magnitude-limited samples of Las Campanas galaxies, and compare
our results to the power spectra derived from other redshift surveys.
In \S~\ref{lumbias} we compute the power spectrum for volume-limited 
samples of Las Campanas galaxies and test for luminosity bias in the survey.
In \S~\ref{cosmo} we compare our power spectrum results against those 
from N-body simulations. We will then focus on the large scale linear power spectrum,
relate our results to the COBE DMR constraints, and compare against the
predictions of several classes of CDM models.
We summarize our results in \S~\ref{ps:conclusions}.
Note that a complementary analysis of the {\em 2-dimensional} LCRS power spectrum has already
been carried out (\cite{lan96}; more on this below), and the derivation of
the closely-related two-point correlation function of LCRS galaxies is described in 
Tucker (1994) and Tucker et al.\ (1996).

\section{Estimating the Power Spectrum} \label{ps:methods}

The power spectrum estimation technique used here has been described by
various authors, in particular see Fisher et al.\ (1993), 
Feldman et al.\ (1994), and Park et al.\ (1994), and we detail the method
below. 
The most important difference is that the LCRS consists of six essentially 
two-dimensional ``slices'', so that we need to account for
``convolution'' effects caused by
the survey geometry in order to calculate the power spectrum properly.
These convolution effects are also evaluated below.

\subsection{Methods} \label{submethods}

Given a galaxy density field $\rho({\bf r})$ with density contrast
$\delta({\bf r}) = [\rho({\bf r}) - \bar\rho] / \bar\rho$, where $\bar\rho$
is mean galaxy density, we have the two-point correlation function
\begin{equation}
\xi({\bf r}) = \langle \delta({\bf r}') \delta({\bf r}' + {\bf r}) \rangle
_{{\bf r}'}
\end{equation}
and the power spectrum 
\begin{equation} \label{eqpk}
P({\bf k}) = \int d^3{\bf r}\ \xi({\bf r}) e^{i{\bf k} \cdot {\bf r}}\ .
\end{equation}
With this definition $P({\bf k})$ is a power spectral density with units
of volume. Assuming isotropy we have $\xi({\bf r}) = \xi(r=|{\bf r}|)$
and $P({\bf k}) = P(k=|{\bf k}|)$.

Now, suppose we have a survey of volume $V$, with $N$ observed galaxies 
at positions 
${\bf r}_i$. The survey's selection function $s({\bf r})$ is defined to
be the fraction of galaxies at position ${\bf r}$ expected to be observable 
by the survey,
given the survey's photometric and other selection criteria. Also, let
$w({\bf r}) \equiv 1 / s({\bf r})$. 
We can then form an estimator $\hat{\delta}({\bf r})$ for the true 
density contrast $\delta({\bf r})$ by
\begin{equation} 
\hat{\delta}({\bf r}) = \frac{1}{\sum^{N}_{i=1} w({\bf r}_i) / V} 
                        \sum^{N}_{i=1} w({\bf r}_i) 
\delta^{(3)}({\bf r}-{\bf r}_i) - 1\ ,
\end{equation}
where $\delta^{(3)}$ is the Dirac delta function and where we have
estimated the mean density $\bar \rho$ by $\sum^{N}_{i=1} w({\bf r}_i) / V$. 
Note that we will use
a circumflex $\hat{}$ to denote quantities estimated from the observations.
We next take the Fourier transform
\begin{eqnarray}
\hat{\delta}({\bf k}) & \equiv & \frac{1}{V} \int d^3{\bf r}\
\hat{\delta}({\bf r}) e^{i{\bf k} \cdot {\bf r}} \\
& = & \frac{1}{\sum^{N}_{i=1} w({\bf r}_i)} 
      \sum^{N}_{i=1} w({\bf r}_i) e^{i{\bf k} 
\cdot {\bf r}_i} - W({\bf k})\ , \label{eqsum}
\end{eqnarray}
where $W({\bf k})$ is the survey window function
\begin{equation} \label{eqwk}
W({\bf k}) \equiv \frac{1}{V} \int_V d^3{\bf r}\ e^{i{\bf k} \cdot {\bf r}}\ .
\end{equation}
We can convert the sum over galaxies in equation~(\ref{eqsum}) to a sum over
infinitesimal cells distributed throughout the survey volume, as in 
Peebles (1980, \S\S~36 and 41), where each cell $i$ holds 
$n_i =$ 1 or 0 galaxies. Then the expectation value of
$|\hat{\delta}({\bf k})|^2$ may be written
\begin{eqnarray} \label{eqdk}
\langle |\hat{\delta}({\bf k})|^2 \rangle  & = &
\frac{1}{(\bar\rho V)^2} \sum_i \sum_j \langle n_i n_j 
\rangle w({\bf r}_i) w({\bf r}_j)
e^{i{\bf k} \cdot ({\bf r}_i - {\bf r}_j)} \\ \nonumber
& + & |W({\bf k})|^2 \\  \nonumber
& - & \frac{1}{\bar\rho V} 
      \sum_i \langle n_i \rangle w({\bf r}_i) 
[e^{i{\bf k} \cdot {\bf r}_i} W^*({\bf k}) + 
e^{-i{\bf k} \cdot {\bf r}_i} W({\bf k})]\ .
\end{eqnarray}
Using equations (\ref{eqpk}) and (\ref{eqwk}), and the relations
\begin{equation}
\langle n_i n_j \rangle = 
\left\{
\begin{array}{lr}
\bar{\rho}^2 [1 + \xi(|{\bf r}_i-{\bf r}_j|)]
s({\bf r}_i) s({\bf r}_j) dV_i dV_j & i \neq j \\
\bar\rho s({\bf r}_i) dV_i & i = j
\end{array}\ ,
\right.
\end{equation}
we can rewrite equation~(\ref{eqdk}) as
\begin{equation}
\langle |\hat{\delta}({\bf k})|^2 \rangle = 
\frac{\tilde{P}({\bf k})}{V} + \langle S \rangle\ ,
\end{equation}
which is the sum of the true power spectrum convolved with the survey window
function,
\begin{equation} \label{eqpkconv}
\tilde{P}({\bf k}) \equiv \frac{V}{(2 \pi)^3} 
\int d^3{\bf k'}\ P(k') |W({\bf k}-{\bf k'})|^2\ ,
\end{equation}
and the expectation value of the shot noise arising from sampling a finite 
sample of galaxies,
\begin{equation}
\langle S \rangle \equiv \frac{1}{\bar\rho V^2} \int d^3{\bf r}\ w({\bf r})\ .
\end{equation}
Note that we are using a tilde $\tilde{}$ to denote the 
{\em convolved} power spectrum.
For a given set of galaxies the shot noise $S$ is given exactly by the $i = j$
terms in the first sum of equation~(\ref{eqdk}),
\begin{equation}
S = \frac{1}{[\sum^{N}_{i=1} w({\bf r}_i)]^2} \sum^N_{i=1} w^2({\bf r}_i)\ ,
\end{equation}
and our first estimate of the observed power spectrum, convolved with the
survey window function, is thus
\begin{equation} 
\hat{\tilde{P}}(k) = V \int \frac{d\Omega_{\bf k}}{4 \pi}\ 
[|\hat{\delta}({\bf k})|^2 - S]\ ,
\end{equation}
where we average over different wavevectors {\bf k} at fixed magnitude
$|{\bf k}| = k$. 
However, because we do not know the true mean density beforehand, but must
estimate it from the survey itself, we will 
underestimate the
power spectrum on scales comparable to the survey size by an amount
$\tilde{P}(0) |W({\bf k})|^2$ (\cite{pea91}). It turns out that for our
survey geometry, $\tilde{P}(k) \approx {\rm constant} \approx \tilde{P}(0)$
on large scales (see the next subsection). Given an initial estimate 
for $\tilde{P}(0)$, we can correct for the underestimate by adding back the 
$\tilde{P}(0) |W({\bf k})|^2$ term,
so that our final estimate for the observed convolved power spectrum is
\begin{equation} \label{eqpkest}
\hat{\tilde P}(k) = \int \frac{d\Omega_{\bf k}}{4 \pi}\ 
  \left(  
    V \left[
      |\hat{\delta}({\bf k})|^2 - S
    \right] + \tilde{P}(0) |W({\bf k})|^2 
  \right) \ .
\end{equation}
We limit our power spectrum computations to scales for which this correction
is $\lesssim 10\%$, which corresponds to $\lambda \lesssim 400 \ h^{-1}$~Mpc
for the LCRS.
Now, the expectation value of $\hat{\tilde P}(k)$ is, using 
equation~(\ref{eqpkconv}),
\begin{eqnarray}
\langle \hat{\tilde P}(k) \rangle & = &
\int \frac{d\Omega_{\bf k}}{4 \pi} \tilde{P}({\bf k}) \\
& = & \int dk'\ k'^2 P(k') K(k,k')\ ,  \label{eqpconvint}
\end{eqnarray}
where 
\begin{equation} \label{eqkerndef}
K(k,k') \equiv \frac{V}{(2 \pi)^3} 
\int \frac{d\Omega_{\bf k}}{4 \pi} \int d\Omega_{\bf k'}
|W({\bf k}-{\bf k}')|^2\ .
\end{equation}
The effect of the convolution integral (\ref{eqpconvint}) for our survey
will be illustrated in the next subsection.
Because convolution effects are large for our survey, it will be 
convenient to deconvolve the observed $\hat{\tilde P}(k)$ to recover $P(k)$.
We will use the iterative method due to Lucy (1974), which has been 
applied in a similar context by Baugh and Efstathiou (1993) to recover the 
spatial power spectrum from 
the angular correlation function. Given the observed convolved power 
$\hat{\tilde P}$ at a set of wavenumbers $k_j$, we can find the true power
$P$ at the wavenumbers $k_i$ by starting with initial guesses $P^0(k_i)$,
and then computing new estimates $P^n(k_i)$ by iteration
\begin{equation}
P^{n+1}(k_i) = P^n(k_i) \frac{\sum_j [\hat{\tilde P}(k_j) / \tilde{P}^n(k_j)] 
K(k_i,k_j) \Delta k}{\sum_j K(k_i,k_j) \Delta k}\ ,
\end{equation}
where
\begin{equation}
\tilde{P}^n(k_j) = \sum_l k^2_l P^n(k_l) K(k_j,k_l) \Delta k \ .
\end{equation}
We also apply a smoothing procedure
\begin{equation}
P^n(k_i) \rightarrow 0.5 P^n(k_i) 
        + 0.25 \left[ P^n(k_{i-1}) + P^n(k_{i+1}) \right]
\end{equation}
in order to reduce noise as the $P^n(k_i)$ converge to a final solution.
On the largest scales $k \lesssim 0.03\ h$~Mpc$^{-1}$,
the smoothing also reduces the sensitivity of the solution to the choice
of initial guess $P^0(k_i)$ (we adopt $P^0(k_i) = $ constant).
Because the deconvolution procedure is not too well 
constrained on large scales, and because the smoothing procedure is somewhat
arbitrary, we will use the deconvolved power spectra
only for qualitative comparisons. The computation of the convolved power
spectrum from a given true power spectrum is more straightforward, and our
quantitative analyses will focus on the convolved power spectrum. 

\subsection{Test of the Methods on N-Body Simulations} \label{subntest}

We check our power spectrum estimation methods on an N-body simulation
kindly provided by Changbom Park. The simulation is an open cold dark matter
model (denoted ODM), with $\Omega_0 = 0.4$ and $h = 0.5$, whose power spectrum
has been found to be a good match to that of the Center for
Astrophysics (CfA) Redshift Survey (\cite{par94}). The model
is unbiased, that is, $b = 1$, where $b$ is the ratio of galaxy to mass 
density fluctuations. The model is normalized so that 
$\sigma_8 = 1$, where $\sigma_8$ is the rms galaxy density fluctuation in 
a sphere of radius 8\ $h^{-1}$~Mpc. The ODM simulation contains $240^3$ 
particles and was computed using a particle-mesh code (\cite{hoc81}) on 
a $480^3$ mesh with corresponding physical comoving volume 
$(576\ h^{-1}$~Mpc$)^3$.

We draw from the ODM simulation a set of ``mock'' redshift surveys using the
geometric and photometric selection criteria of our actual survey 
(see \S~\ref{pslcrs}). The ODM ``galaxies'' are assigned absolute magnitudes 
using the luminosity function derived for the LCRS 
(\cite{lin96lf}).
The velocities of the ODM galaxies are included so we
measure the power spectrum in redshift space. For each mock survey we
rederive the selection function and then compute the power spectrum using
the methods detailed in \S~\ref{submethods}.
Now, even though the ODM simulation box is quite large, the full LCRS still
only fits along a body diagonal of the ODM box. We can however fit either the
Northern or Southern halves of the survey into the box fairly readily, and
we have drawn a series of Northern and Southern LCRS mock surveys. To make
a reasonably large number of {\em whole}-LCRS mock surveys, we chose to combine 
a Northern and a Southern mock survey even though these two mock surveys 
are not {\em originally} oriented relative to each other in the 
simulation box as the real Northern and Southern surveys are.
Figure~\ref{figodmsim} shows the average power spectrum measured  
from 30 whole-LCRS mock surveys. We show both the directly measured, or
convolved, power, as well as the deconvolved power of the mock
surveys. The errors plotted are the standard deviations of the mean
(sdom) of the 30 surveys.
Also plotted in the figure are the true unconvolved power spectrum of the ODM
simulation and the ODM power spectrum after convolution with our 
whole-survey window function (using equations~[\ref{eqpconvint}] 
and [\ref{eqkerndef}]). 
There is good agreement between the
true ODM power spectrum and the average power spectrum of
the mock surveys over the range of scales
$\lambda = 5 - 400\ h^{-1}$~Mpc that we examine, showing that
we can correctly account for the geometry and 
selection criteria of the LCRS in our measurement and deconvolution of
the mock-survey power spectra.

Note from Figure~\ref{figodmsim} that the main effect of the convolution
by the window function of our survey is that the convolved power spectrum
lies below the unconvolved power spectrum over most of the scales probed.
Also, note that on the largest scales, wavelengths 
$\lambda > 100\ h^{-1}$~Mpc, the
convolved power spectrum is flat whereas the true power spectrum may peak and
begin to turn over. (This is seen more clearly for the power spectrum
of Figure~\ref{fignsfit}.) In Figure~\ref{figkernel} we illustrate the convolution
effect in more detail by plotting the integrand 
$k'^2 P(k') K(k,k')$ of the convolution integral~(\ref{eqpconvint}) for
the ODM model at several values of $k$.
This integrand shows the 
contribution to the convolved power $\tilde{P}$ at $k$ from the true power 
$P$ at different $k'$. For a 3-dimensional
survey geometry, such as a sphere or cube, at scales $\lambda = 2 \pi / k$ 
small compared to the survey size, we would expect this
integrand to be sharply peaked at $k' = k$; that is, convolution effects would
not be important and the measured convolved power would be very close to the 
true power. However, the Las Campanas survey has essentially a two-dimensional
geometry: it is large in two dimensions but thin along the third
(cf. Figure~\ref{figslices}). Convolution
effects can be noticeably important at scales smaller than the survey size, 
and ``aliasing'' occurs: power at $k' \neq k$ makes a significant 
contribution to the convolved power $\tilde{P}$ at $k$, leading to the effects
noticed from Figure~\ref{figodmsim}.
As seen in Figure~\ref{figkernel}, for $k \lesssim 0.1\ h$~Mpc$^{-1}$, 
the integrand is broad and somewhat complicated
in shape, including contributions from a large range of $k'$. The situation
improves for smaller scales and larger $k$, as the integrand becomes 
more and more sharply peaked at $k' = k$. Nevertheless, though these 
convolution effects are complicated in detail, they can be computed 
and accounted for in the power spectrum analysis, as we saw earlier.

\section{Power Spectrum of LCRS Galaxies} \label{pslcg}

\subsection{Power Spectrum of LCRS Samples} \label{pslcrs}

We compute the power spectrum in redshift
space, but first convert our heliocentric redshifts to comoving distances
$r$, assuming $\Omega_0 = 1$. Using a different value for $\Omega_0$
does not make much difference; for example, $r$ changes by 2\% at $z = 0.1$
if we instead use $\Omega_0 = 0.2$.
We first consider the power spectrum for magnitude-limited samples of the 
LCRS. We thus need to weight by the inverse of the selection function,
but we note that this weighting becomes large when the selection function 
becomes small at large 
distances, where only the intrinsically brightest and rarest galaxies are 
observed. Also, because
of the surface brightness cuts imposed on the sample, we do not probe
the galaxy luminosity function well at the faint end (\cite{lin96lf}), but
these intrinsically faint galaxies could make a large contribution in the 
weighting. To reduce the sensitivity of our power spectrum results to these 
potential problems,
we make additional restrictions on the redshifts and absolute magnitudes
of the galaxies used in our clustering analyses: 
10000 km~s$^{-1} < cz <$ 45000 km~s$^{-1}$ (or $98 < r < 405\ h^{-1}$~Mpc)
and $M_1 \equiv -22.5 < M - 5 \log h < -18.5 \equiv M_2$.
These limits keep the selection function $s(cz) \gtrsim 0.05$, and
result in a total sample of 19305 galaxies.
The weight function $w$ used in equation~(\ref{eqsum})
for galaxy $i$ of the survey is given by
\begin{equation} \label{eqweight}
w({\bf r}_i) = \frac{W_i}{s({\bf r}_i)}\ ,
\end{equation}
where $W_i$ is a weighting factor which accounts for variable 
field-to-field sampling fractions and incompletenesses as a function of 
apparent magnitude and central surface brightness (details in \cite{lin96lf}), 
and $s({\bf r}_i)$ is the selection function appropriate to the survey field 
containing galaxy $i$,
\begin{equation} \label{eqphi}
s({\bf r}_i) = 
   \int^{min[M_2({\bf r}_i),M_2]}_{max[M_1({\bf r}_i),M_1]} \phi(M) dM
   \left/
     \int^{M_2}_{M_1} \phi(M) dM
   \right. \ .
\end{equation}
Here $M_1({\bf r}_i)$ denotes the minimum and $M_2({\bf r}_i)$ 
the maximum absolute magnitude
observable at the distance $r$, given the apparent magnitude limits of the 
field of galaxy $i$, and $\phi(M)$ is the LCRS luminosity function.
For $\phi(M)$ we use a Schechter luminosity function (\cite{schech76})
with parameters $M^* = -20.3 + 5 \log h$ and $\alpha = -0.7$;
details concerning the derivation of the luminosity function are given in
Lin et al. (1996). 

Figure~\ref{fig415}(a) shows the observed convolved $\hat{\tilde{P}}(k)$ for 
the full LCRS
sample defined above, computed using the methods detailed in \S~\ref{ps:methods}.
We have also split the full sample into four subsamples by dividing each of the
Northern and Southern data subsets in half by RA; see Table~\ref{ps:tabsamps}
for details. The individual
$\hat{\tilde{P}}(k)$ for the four subsamples are also plotted in 
Figure~\ref{fig415}. We use the standard deviations of the mean of
$\hat{\tilde{P}}(k)$ for the subsamples, or the $1 \sigma$ errors from
the whole-LCRS ODM mock surveys of \S~\ref{subntest}, whichever is larger, 
to estimate the $1 \sigma$ errors for the full sample. We also note
that though we average $\hat{\tilde{P}}({\bf k})$ over different 
directions of $\vert {\bf k} \vert$
at fixed $\vert {\bf k} \vert = k$ in equation~(\ref{eqpkest}), results 
from the mock surveys show that the errors depend only weakly on direction.
The convolution 
effects are quite similar for all five samples shown in 
Figure~\ref{fig415}(a), so that the convolved power spectra may be directly
compared. There is good agreement among the four subsamples for wavelengths
$\lambda \lesssim 100\ h^{-1}$~Mpc. On larger scales, aside from sampling 
fluctuations, uncertainties in the survey's selection function and mean density
will also contribute to the differences among the samples.
We have also divided the data into 50-fiber and 112-fiber
subsets (see Table~\ref{ps:tabsamps}), 
and computed the convolved power spectrum for
each subsample, with the results plotted in Figure~\ref{fig415}(b). Again,
the convolution effects for these two subsets are similar so we directly
compare them.
There are no conspicuous systematic differences between the 
power spectra computed from 50- vs.\ 112-fiber data, giving us confidence
that the power spectrum is not sensitive to the different sampling and
photometric selection criteria involved. The two subsamples agree well 
for $\lambda \lesssim 150\ h^{-1}$~Mpc. 

We can also check the consequences on the power spectrum arising from
two geometric selection effects of the survey. First, recall from 
\S~\ref{intro} that an instrumental constraint prevents two galaxy
fibers from approaching closer than $55\arcsec$ on the sky. In the ODM
mock surveys we have actually applied this restriction and our
previous results indicate apparently little effect on the power
spectrum. More explicitly, we can assign fake velocities to those
unobserved objects which lie within $55\arcsec$ of an observed survey galaxy, and
then recompute the power spectrum. 
Some fraction of these pairs will be physically associated,
and we assign velocities to the unobserved objects according to the distribution
of pairwise velocity differences for observed survey galaxy pairs with
angular separations between $1\arcmin$ and $2\arcmin$. For this
purpose we use a very generous upper bound of 5000 km~s$^{-1}$ for the 
velocity difference, corresponding to 60\% of the observed velocity
difference distribution.
The remaining 40\% of pairs are assumed to be chance superpositions,
for which we assign velocities to the unobserved objects 
according to the overall redshift histogram of
the particular survey field containing that pair.
Note that this procedure adds an average of 1351 galaxies to
the full LCRS sample used for power spectrum computation.
Figure~\ref{figgeo}(a) plots the full range of the power spectrum computed from 10
different sets of such velocity assignments, and confirms that the full-sample
LCRS power spectrum is indeed very little affected over the scales probed ($<
10\%$ difference).
Second, as described in Shectman et al.\ (1995), the spectroscopic
success rate is lower (though still $> 90\%$) at the corners of the 
spectroscopic fields. We can approximately compensate for this effect
by calculating a ``neighborhood'' sampling fraction within an area of
radius $1000\arcsec$ centered on each galaxy, and then applying this
additional geometric weighting in (\ref{eqweight}). The choice of 
regions of $1000\arcsec$ radius to define the weights came from trial
and error efforts to best reproduce the angular pair distribution of the survey
photometric catalog, by appropriately weighting the pair distribution
of the spectroscopic sample.
Figure~\ref{figgeo}(b) shows that applying this correction also has
little effect on the LCRS power spectrum. For simplicity, since both the above
geometric effects are small, we will neglect them in the 
remainder of this paper.

We next use the Lucy deconvolution procedures outlined in \S~\ref{ps:methods} 
to obtain estimates of the deconvolved
$\hat{P}(k)$ for the full sample and the four subsamples, with the results for
the full sample displayed in Figure~\ref{fignsfit}. 
Here we again estimate errors using the
sdom of the four subsamples, or the $1 \sigma$ errors from the 
deconvolved power spectra of the whole-LCRS ODM mock surveys, whichever is
larger.
Besides using Lucy deconvolution, we have also convolved a convenient model
for the power spectrum and fit to the observed LCRS 
full sample $\hat{\tilde{P}}(k)$:
\begin{equation} \label{eqpk4}
P(k) = \frac{2 \pi^2}{k^3} 
       \frac{(k/k_0)^{n_0}}{1 + (k_c/k)^{n_c}} \ .
\end{equation}
The model was motivated by a similar fitting formula applied to the APM
data by Peacock (1991) (see also \cite{fel94}). The model has four free
parameters ($k_0, n_0, k_c, n_c$), and essentially it behaves as a power
law with slope $n_0-3$ at small scales (large $k$), changing to another
power law with slope $n_0-3+n_c$ at large scales (small $k$) 
(we take $n_0, n_c > 0$). The transition
occurs near wavenumber $k_c$, and there remains an overall normalization
that can be more meaningfully expressed in terms of $\sigma_8$. We
find
\begin{eqnarray}
  k_c & = & (0.06 \pm 0.01)\ h\ {\rm Mpc}^{-1} \nonumber \\
 P(k) \propto k^{n_0-3} & = & k^{-1.8 \pm 0.1}\ , \
  2\pi / k = 5 - 30\ h^{-1} \ {\rm Mpc}
  \nonumber \\
 P(k) \propto k^{n_0-3+n_c} & = & k^{1 \pm 1}\ , \
  2\pi / k \approx 200 - 400\ h^{-1} \ {\rm Mpc} \\
  k_0 & = & (0.17 \pm 0.01)\ h\ {\rm Mpc}^{-1} \nonumber \\
  \sigma_8 & = & 1.0 \pm 0.1\ {\rm (redshift\ space)} \nonumber
\end{eqnarray}
We plot the fit in Figure~\ref{fignsfit}. Note that for 
$\lambda \lesssim 100 \ h^{-1}$~Mpc, the power spectrum is well-determined.
There is an apparent peak at about $2\pi / k_c = 100 \ h^{-1}$~Mpc, but
on larger scales the form of $P(k)$ is not too well determined, with our fit
power law slope consistent with a flat to declining $P(k)$ with decreasing $k$.
We want to emphasize that for $\lambda \gtrsim 100 \ h^{-1}$~Mpc, the
errors are unavoidably large and we do not claim detection of a turnover
in the power spectrum, only that our results suggest that $P(k)$ is at least
flattening out on these largest scales. Improved power spectrum estimation
techniques, as recently described by Tegmark (1995) or by Vogeley 
\& Szalay (1996), may prove to be helpful in reducing the errors on
the large-scale power spectrum, though application of these methods is
beyond the scope of the present paper.

As we mentioned in \S~\ref{intro}, a complementary analysis of the 2D
LCRS power spectrum may be found in Landy et al.\ (1996). 
The wavevectors in the 2D treatment are
aligned along the ``planes'' of each of the six slices, rather than
averaged over the whole sphere as they are in this paper. 
The 2D analysis thus emphasizes those spatial directions for which the 
survey geometry is best suited on large scales. This renders the 
2D analysis less sensitive to the 
presence of aliased power and therefore more sensitive to the presence of 
features in the power spectrum on large scales.
In particular, a striking peak at $\lambda \approx 100 \ h^{-1}$~Mpc is
seen in the 2D power spectrum, and it is a signature of the clear
pattern of over- and under-dense regions (e.g., walls and voids) in
the LCRS galaxy distribution (Figure~\ref{figslices}). A different
``core-sampling'' analysis applied to the LCRS by Doroshkevich et al.\ (1996) 
also indicates that $100 \ h^{-1}$~Mpc is approximately the typical
separation between large sheet-like structures in the galaxy distribution.
These results are also reminiscent of a similar peak at $128 \ h^{-1}$~Mpc 
in the 1D power spectrum of deep pencil beam surveys 
analyzed by Broadhurst et al.\ (1990). 
A peak at $\lambda \approx 100 \ h^{-1}$~Mpc
is also suggested in the 3D power spectrum fit above, though it 
does not stand out in Figure~\ref{fignsfit}, both because of the
logarithmic y-axis scaling and because of the suppression due to aliasing as
mentioned above. The focus of the 2D paper is on the $100 \ h^{-1}$~Mpc scale
power, and the reader is
referred to Landy et al.\ (1996) for the full details. 
On the other hand, the current analysis has the broader goal of relating
the full LCRS power spectrum, in the more usual 3D context,
to results from previous surveys, to COBE constraints, and to predictions of 
cosmological models.

\subsection{Comparison to Other Surveys} \label{otherz}

We proceed now to compare the LCRS power spectrum to that determined
from several other large galaxy
redshift surveys. First, Figure~\ref{figcfa} shows the comparison to the power
spectra of three samples drawn from the SSRS2 and CfA2 redshift surveys: 
two volume-limited
samples SSRS2+CfA2 101 ($r < 101.05 \ h^{-1}$ Mpc, $M_B < -19.7 + 5
\log h$) and 
SSRS2+CfA2 130 ($r < 130.0 \ h^{-1}$ Mpc, $M_B < -20.3 + 5 \log h$; \cite{daC94}), 
and one 
apparent-magnitude-limited sample CfA 101m 
($r < 101.05 \ h^{-1}$ Mpc, $m_{B(0)} < 15.5$; \cite{par94}). 
Now, while the Las Campanas galaxies
have been selected from photometry in a red wavelength band, the SSRS2 and 
CfA2 galaxies were selected in the blue. As a rough guide to compare
the red and blue magnitudes, we note
that $M^*_{LCRS} = -20.3$ (\cite{lin96lf}), while
$M^*_{CfA2} = -18.8$ (\cite{marzke94}) and $M^*_{SSRS2} = -19.5$ (\cite{daC94a}).
Nonetheless, the LCRS $P(k)$ is in remarkably good agreement with the $P(k)$ of 
the magnitude-limited CfA101m and volume-limited SSRS2+CfA2 101 samples. 
There are some differences, though they are not at
high significance given the errors. Specifically, the LCRS power
spectrum is lower than SSRS2+CfA2 101
on intermediate scales $\lambda \sim 20-40\ h^{-1}$~Mpc,
and it is also low relative to these other two samples
on large scales $\lambda \gtrsim 150\ h^{-1}$ Mpc. In contrast, the power 
spectrum of the volume-limited SSRS2+CfA2 130 sample is higher than that
of the LCRS on {\em all} scales examined. 
Most simply, these $P(k)$ differences may just 
be due to cosmic variance that arises from the different volumes being
sampled in the three surveys.
Alternatively, luminosity bias might help reconcile some of the differences
we see, in the sense that intrinsically brighter galaxies are more strongly
clustered. We note that the two volume-limited SSRS2+CfA2 samples are restricted
to galaxies brighter than $M^*_{CfA2}$ and $M^*_{SSRS2}$, whereas
the LCRS sample contains galaxies between approximately $M^*_{LCRS} + 2$ and
$M^*_{LCRS} - 2$.
The magnitude-limited CfA101m sample does contain galaxies both
brighter and fainter than $M^*_{CfA2}$, and seems the best match to 
the LCRS. Park et al.\ (1994) find evidence
for luminosity bias in the CfA survey, and we explore this possibility
for the LCRS in \S~\ref{lumbias}. As another alternative, the differences 
between the two surveys may reflect intrinsic clustering differences of
galaxies selected in different wavelength bands, red for the LCRS 
vs.\ blue for the other two surveys, analogous to the effect observed 
for CfA galaxies vs.\ infrared-selected IRAS galaxies (see below). 
The galaxy sampling and surface brightness 
selection criteria of the LCRS are unlikely to produce a systematic bias in
our power spectrum estimates over the
scales probed, given the results of the test done in \S~\ref{subntest}
and the agreement of the 50- and 112-fiber results in Figure~\ref{fig415}(b). 

We next compare the LCRS $P(k)$ to that found from two redshift surveys of
IRAS galaxies: the 1.2 Jy survey (60 $\mu$m flux $<$ 1.2
Jy; \cite{fis93}) and the deeper but more sparsely sampled QDOT sample 
(60 $\mu$m flux $<$ 0.6 Jy; \cite{fel94}). 
As seen in Figure~\ref{figiras}, the IRAS 1.2 Jy $P(k)$ 
is lower than that of the LCRS over all scales probed, particularly for
$\lambda < 30\ h^{-1}$~Mpc. Fisher et al.\ (1993) have shown
that the 1.2 Jy $P(k)$ is very similar in {\it shape} to the CfA 101 $P(k)$,
but that the overall clustering amplitude for IRAS galaxies is lower.
In redshift space $\sigma_8({\rm 1.2\ Jy}) = 0.8$, 
$\sigma_8({\rm CfA 101}) = 1.1$, while the LCRS is in between, with
$\sigma_8({\rm LCRS}) = 1.0$. The clustering strength seen in the 1.2 Jy
survey may reflect the inherently weaker clustering amplitude of the IRAS galaxy
population, which tend to be spirals that avoid rich clusters 
(e.g. \cite{str92}). We have selected from the LCRS a sample
of emission galaxies, with [OII] $\lambda$3727 equivalent widths 
$W_{\lambda} \geq 5$~\AA,
which should also tend to be spirals, as
well as a sample of galaxies with $W_{\lambda} < 5$~\AA, which 
should include more ellipticals (see Table~\ref{ps:tabsamps}). 
We compare the relative convolved power spectra of the whole, 
emission, and non-emission LCRS
samples in Figure~\ref{figem5}. We find that for scales
$\lambda \approx 10 - 100 \ h^{-1}$~Mpc, the emission sample
is about 30\% more weakly clustered than the full sample, while the
non-emission sample is about 25\% more strongly clustered than the full
sample, and these results are roughly independent of scale. 
The same trends persist up to $\lambda \approx 300 \ h^{-1}$~Mpc,
though the errors become larger for $\lambda \gtrsim 100 \ h^{-1}$~Mpc
(here again the errors have been estimated from the variation among the 
four LCRS subsamples divided by hemisphere and right ascension).
Though the difference is not as great
as for IRAS vs.\ LCRS galaxies, our results are consistent with an intrinsically
weaker clustering strength of spiral and emission-line galaxies on scales
$\lambda \lesssim 100 \ h^{-1}$~Mpc. 
Galaxies with early morphological types have been observed to be more 
strongly clustered than
late types in a fair number of previous smaller 
samples, exploring generally smaller clustering scales 
(e.g., \cite{dav76}, \cite{hjmo92}, \cite{loveday95b}).
Our current results are thus in qualitative agreement, though the lack
of one-to-one correspondence between morphologies and emission 
strengths vitiates a more quantitative comparison.

We note next from Figure~\ref{figiras}
that the LCRS and QDOT results are in somewhat better agreement, except 
that the QDOT slope is steeper, in
the sense of having more relative power for scales 
$\lambda \sim 100-200\ h^{-1}$~Mpc compared to scales 
$\lambda \sim 50\ h^{-1}$~Mpc. Feldman et al.\ (1993) also point out a
turnover in the QDOT $P(k)$ for $\lambda \gtrsim 150\ h^{-1}$~Mpc.
The LCRS $P(k)$ shows an apparent peak at roughly similar scales,
$\lambda \approx 100\ h^{-1}$~Mpc, but the constraint on the slope of
the LCRS power spectrum on larger scales is not strong, as it can 
accommodate both declining to flat power spectra for $\lambda \gtrsim 200\
h^{-1}$~Mpc (\S~\ref{pslcrs}).
Note finally that the QDOT and 1.2 Jy results differ somewhat for 
$\lambda > 100\ h^{-1}$~Mpc. Feldman et al.\ (1994) attribute this to sampling
fluctuations rather than to luminosity bias among IRAS galaxies
(see also \cite{man95}). The sampling fluctuation argument is also
indicated by the analysis of Tadros \& Efstathiou (1995), who have
calculated the power spectrum of the combined 1.2 Jy and QDOT samples.

Overall, for scales $\lambda \gtrsim 100\ h^{-1}$~Mpc, despite some apparent
differences, the power spectra from the various surveys are broadly 
consistent with each other given
the large errors common to all the surveys. On smaller scales, the power
spectra of the different surveys show similar shapes, but there are 
differences in amplitude which may reflect intrinsic variations in 
clustering strength for galaxies of different luminosities or types. We will
explore the dependence of clustering on luminosity for LCRS galaxies 
next. Nonetheless, it is noteworthy that the {\em independent}, 
{\em optically}-selected surveys LCRS and SSRS2+CfA2, which also provide the 
{\em largest} galaxy samples, do in fact agree well over the range of scales probed.

\section{Luminosity Bias in the LCRS} \label{lumbias}

As we saw in \S~\ref{otherz} and Figure~\ref{figcfa}, 
the SSRS2+CfA2 130 sample ($M_B < -20.3$), 
which contains intrinsically brighter galaxies than the SSRS2+CfA2 101 sample 
($M_B < -19.7$), has a power spectrum amplitude $\approx 40\%$
greater than that of SSRS2+CfA2 101 (\cite{par94}; \cite{daC94}).
Motivated by this evidence of luminosity bias,
here expressed as the variation of 
power spectrum amplitude with galaxy luminosity in the SSRS2 and CfA2 surveys, 
we would like to test for evidence of the same in the LCRS. 
Recall that these two samples are volume-limited, with galaxies brighter than 
$M^*_{CfA2} = -18.8$ and $M^*_{SSRS2} = -19.5$, 
whereas our earlier magnitude-limited LCRS sample contains galaxies both 
brighter and fainter than $M^*_{LCRS} = -20.3$.
Volume-limited samples, defined
by both a lower and an upper bound on absolute magnitude, are a more natural
choice over apparent-magnitude-limited samples for the purpose of examining
how the power spectrum changes as a function of absolute luminosity. Also,
for volume-limited samples the selection function is simpler to deal with.
However, the $1.3$-mag band in the isophotal magnitude limits
of the 50-fiber data restricts drawing volume-limited samples to 
inconveniently small
distance and absolute magnitude ranges, so we will confine the analysis to
the 112-fiber data, which have wider $2.7$-mag isophotal limits. 
Within the same redshift limits 10000 km~s$^{-1} < cz <$ 45000
km~s$^{-1}$  as applied to the
earlier apparent-magnitude-limited full LCRS sample, we form seven 
volume-limited samples,
with absolute magnitude limits $-18.5 > M > -19.5$, $-19.0 > M > -20.0$, 
$\ldots$, $-21.5 > M > -22.5$.
Table~\ref{tabvolsamps} summarizes details of these samples. 
In the power
spectrum calculation, we set the weight function for galaxy $i$ to 
$w({\bf r}_i) = W_i$ (see \S~\ref{pslcrs}).

One complication that we note from Table~\ref{tabvolsamps} is that the 
different volume-limited samples do not necessarily overlap much in space 
with one another, and one might wonder whether any differences seen among 
the samples may be due to sampling fluctuations rather than to actual
luminosity bias. In order to better isolate the luminosity dependence, we 
adopt a procedure similar to that applied by Hamilton (1988) in examining the
variation of the correlation function amplitude with galaxy luminosity in 
the CfA1 survey. We divide each volume-limited sample in half by absolute
magnitude (i.e., the $-19.0 > M > -20.0$ would be split into two samples with
$-19.0 > M > -19.5$ and $-19.5 > M > -20.0$), calculate the ratio of the 
power spectra of the brighter and fainter subsets (which are located in 
the same volume of space), and finally multiply (or divide) these ratios in 
successive volume-limited samples in order to determine the relative
clustering amplitudes as a function of luminosity. We arbitrarily set
the clustering amplitude of the $-20.0 > M > -20.5$ sample, which contains
$M^*_{LCRS} = -20.3$, to unity. We apply this procedure to the whole-survey
volume-limited samples, and estimate errors from the variation among 
volume-limited samples from the four subsets of the whole survey divided
by hemisphere and right ascension (recall Figure~\ref{fig415}(a)). We
also average our results over four roughly equal bins in $\log \lambda$,
as shown in Figure~\ref{figpowrat}: $\lambda = 5-15$, $20-50$, 
$60-150$, and $200-400 \ h^{-1}$~Mpc. We note first from the figure that
results for the bin at $200-400 \ h^{-1}$~Mpc are quite noisy and the 
errors are large enough to be consistent with no variation of clustering 
amplitude with luminosity. We will focus attention on the three bins
in the range from $5-150 \ h^{-1}$~Mpc, where the errors are not so large.
We also find that
the variation of clustering amplitude with absolute magnitude is fairly 
scale-independent over this range of $\lambda$. We see that for
$-18.5 > M > -21.0$ the clustering amplitude changes little with absolute
magnitude. For galaxies
brighter than $M = -21.0$, the clustering amplitude appears to rise with
luminosity; in particular, for $-21.5 > M > -22.0$, the brightest bin
with still reasonably small errors, the clustering amplitude
is $1.6 \pm 0.5$ that at $M^*$. The departures from no luminosity
bias are relatively mild, except in the brightest bin, where 
unfortunately the errors are also large. Nevertheless, there does
appear to be about a 50\% stronger clustering for galaxies brighter than about 
$M^* - 1$ relative to those fainter, in general agreement with the amount 
of luminosity bias seen in the CfA2 and SSRS2 surveys. The trends
seen in Figure~\ref{figpowrat} are similar to analogous results for 
scales $\lesssim 10 \ h^{-1}$~Mpc from the much smaller CfA1 survey 
(\cite{ham88}). Our results are also similar to recent findings from the 
large-volume, but 1-in-20 sparse-sampled
Stromlo-APM survey (\cite{loveday95a}, \cite{loveday95b}), except that we do 
not see a much weaker clustering amplitude for sub-$M^*$ galaxies 
that Loveday et al.\
find. The difference might be due the fact that we do not use galaxies
fainter than about $M^* + 2$ that were included in the analysis of the 
Stromlo-APM data. In addition, recent correlation function analysis of
the SSRS2 sample (\cite{beno96}) indicates a luminosity bias trend
similar to that seen in the LCRS.
We note however that the LCRS provides advantages with its combination of
sample size, sampling density, and 
depth over other redshift surveys, and likely provides the best current 
sample to examine the luminosity bias effect.

The fact that the luminosity bias we see for super-$M^*$ galaxies
persists to fairly large scales argues for a primordial rather than a local
environmental effect as its root cause. Specifically, our results are
qualitatively consistent with the biased galaxy formation scenario, in
which brighter galaxies form out of higher peaks in the underlying 
mass distribution and are therefore more strongly clustered, because such 
higher peaks are {\em naturally} more clustered in models like CDM with
an initially Gaussian field of mass density fluctuations 
(\cite{bar86}; \cite{whi87}). Also, the large-scale
clustering bias seen earlier for our emission and non-emission samples
is also tied to some extent into the luminosity bias issue, as we
have found that our non-emission sample is dominated by super-$M^*$ galaxies
and our emission sample by sub-$M^*$ objects (\cite{lin96lf}). 
If the emission ($\sim$ morphology)
and luminosity bias effects are independent, as evidenced in the
Stromlo-APM survey (\cite{loveday95b}), then some primordial influence
would also be suggested in relation to the galaxy star formation rate, 
for which
[OII] $\lambda$3727 is a rough indicator (\cite{kennicutt92}). In the present paper
we have simply laid out our initial observations of these type-dependent 
clustering differences; we plan a more detailed 
examination of the the effects we have found and
their quantitative implications for biased galaxy formation models.
In addition, note that our present results have been limited to 
redshift-space, and it will be useful, though expectedly more difficult 
and noisy, to factor out the 
redshift distortion effects by examining the clustering in real space, which
we also plan to attempt.

We can however draw some brief conclusions in the context of the
recent semi-analytic galaxy formation models of Kauffmann et al.\
(1995). In particular, their work indicates that the bias factor $b$
should depend only weakly on scale (also \cite{wei95}), consistent
with our observations. Also, their models predict that the amount of
luminosity bias should decrease with increasing $\sigma_8$
of the underlying {\em mass} power spectrum. Specifically, their
Figure~6 indicates that on scales $\sim 20 \ h^{-1}$~Mpc, galaxies
with $M \lesssim M^* + 1$ are more biased relative to those
with $M \lesssim M^* + 3$, by factors of about 1.5 for
$\sigma_{8,mass} = 0.4$ and $\lesssim 1.2$ for $\sigma_{8,mass} = 1$.
For comparison, we can convert the relative power values of
Figure~\ref{figpowrat} into relative bias
values (recall power $\propto b^2$), and then make the appropriate sum
over the LCRS luminosity function in order to determine the average
bias over some absolute magnitude range. We find for scales
$5 - 150 \ h^{-1}$~Mpc that the bias
of $M \lesssim M^*$ galaxies is $< 1.1$ times that of $M \lesssim
M^* + 2$ galaxies (the stronger clustering of bright galaxies is more
than compensated by their reduced numbers in the luminosity function).
Our results are on the weak luminosity bias end of the range of
Kauffmann et al.\ models, and thus
imply that $\sigma_{8,mass}$ should be high, 
$\approx 1$. As this is near the observed {\em galaxy} $\sigma_8$ , 
it also indicates
that LCRS galaxies may be unbiased tracers of the mass. We return to
this point below in \S~\ref{linear}.

\section{Comparisons with Cosmological Models} \label{cosmo}

We now compare the LCRS results against the expectations of
cosmological models. We begin with a brief comparison using two
particular N-body
simulations, which allow us to examine the model power spectra over
both small and large scales. Then we proceed to widen the explorable
parameter space by concentrating on fitting the LCRS power spectrum on
large scales in the linear regime, and compare against the predictions
of several classes of CDM-motivated models. In
combination with COBE and other large scale structure constraints, the
large scale LCRS power spectrum will help us delineate the allowed
parameter space in these cosmological models.

\subsection{Comparison to N-Body Simulations} \label{nbody}

We compare the LCRS power spectrum to that of two N-body models, both kindly
provided by Changbom Park: (1) the ODM model of \S~\ref{subntest},
with $\Omega_0 = 0.4$, $h = 0.5$, and bias factor $b = 1$; and (2)
the CDM1 model, which has $\Omega_0 = 1$, $h = 0.5$, and $b = 1.5$.
The CDM1 model was computed with a particle-mesh code on a $324^3$ mesh, 
physical comoving
volume $(388.8\ h^{-1}$~Mpc$)^3$, and contains $162^3$ CDM particles and
$1,201,320$ biased ``galaxy'' particles, chosen by the biasing scheme of Park 
(1991). Both models are normalized so that the galaxy $\sigma_8 = 1$.

The {\em convolved} redshift-space power spectra of the LCRS and the N-body 
models are plotted
in Figure~\ref{figpnsdm}. 
The three power spectra agree well with each other for
$\lambda < 20\ h^{-1}$~Mpc. On intermediate scales $\lambda \sim 30$ to
$50\ h^{-1}$~Mpc, the CDM1 model matches the LCRS results somewhat better than
the ODM model does.
For wavelengths $\lambda > 60\ h^{-1}$~Mpc,
the LCRS power falls in between the CDM1 and ODM curves, with the LCRS
results closer to that of CDM1 on the largest scales 
$\lambda > 100\ h^{-1}$~Mpc. Though both N-body
models match the LCRS on small scales, neither model provides quite the
right amount of large-scale power: not enough power in the case of CDM1 and 
too much in the case of ODM.

To be more 
quantitative in our comparison, we can use a rank-sum test to see how
probable it is to draw the LCRS power spectrum from the population of N-body
mock-survey power spectra. We choose the rank-sum test as it is simpler 
to apply than a $\chi^2$ test; we need not make any assumptions about 
either the distribution of power at each $k$ or about the correlations 
between the power at different $k$. At each
wavenumber $k$, we first assign each of the 30 ODM mock-survey 
plus 1 LCRS samples a rank $R_{i,k}$ in order of increasing power 
(the index $i$ denotes the sample). 
Then, for each of the 31 samples we
combine the ranks at different $k$ by forming the sum 
$S_{i} = \sum_k R_{i,k}$. Finally we assign each sample an overall rank
$R_{i,sum}$ in order of ascending $S_i$. Doing this results in the LCRS 
receiving a rank $R_{LCRS,sum} = 2$, giving us a two-tailed confidence interval 
of $100 - 2 \times (2 / 31 \times 100)$ = 87\% for rejecting the null 
hypothesis that the LCRS power spectrum is drawn from the population of 
ODM power spectra. This is not very high significance, and it will
be further weakened by the fact that since we only have one ODM simulation box,
we do not sample the full range of variation as we would given an entire 
ensemble of ODM simulation boxes. Nevertheless, the rank-sum test does 
give us a simple quantitative measure of how well the ODM model can match
the LCRS data. Likewise, the rank-sum test applied to 30 CDM1 mock-survey
plus 1 LCRS
samples gives $R_{LCRS,sum} = 10$, where we now assign $R_{i,k}$ in order 
of {\em decreasing} power because the LCRS power is greater than that of 
most of the CDM1 mock surveys. The two-tailed confidence interval for 
rejecting CDM1 is then just 35\%. Repeating the rank-sum tests, but now
just focusing on large scales $k \leq 0.1\ h$ Mpc$^{-1}$, we
find $R_{LCRS,sum} = 1$ for both ODM and CDM1, so now
the rejection probabilities are 94\% in both cases. The match of the
CDM1 model to the LCRS is thus appreciably worse when considering only
large scales, while the match of the ODM model is little changed.
Neither model 
provides a completely satisfactory match to the LCRS power spectrum over all
the scales which we probe. The LCRS power spectrum
lies between that of the two models, and as we will see in the next section,
the LCRS power spectrum on large scales is better matched by a linear 
CDM model with $\Omega_0 h \approx 0.3$.

\subsection{Fitting the Linear Power Spectrum and Comparison to COBE} 
\label{linear}

For any particular structure formation model, the best comparison procedure
with observations is to draw out mock redshift surveys as we did in 
\S~\ref{subntest}. However, it will be useful to fit 
our observations to analytic models in order to explore more of 
cosmological parameter space than permitted by Park's two N-body models.
To do a proper fit to a 
{\em nonlinear, redshift-space} power
spectrum given a {\em linear, real-space} model requires us to account for 
the following: (1) the nonlinear evolution of the power spectrum; (2) the bias 
relating the galaxy fluctuations to the underlying mass fluctuations; and (3) 
the distortions caused by mapping from real to redshift space. Though 
these three effects may be modelled analytically (\cite{pea94};
\cite{jai95}), we will
take a simpler route and concentrate on fitting 
just those observations on large scales in, at least approximately, the 
linear regime. We thus sacrifice the information
we have on small scales for the simplicity of not having to model the 
nonlinear effects (which perhaps are best dealt with via N-body simulations).

To specify our assumptions, note the following relations 
connecting the galaxy power spectrum and the underlying mass power spectrum:
\begin{eqnarray}
P_{galaxy,real-space}(k) & = & f_{b}(k) P_{mass,real-space}(k) \\
P_{galaxy,redshift-space}(k) & = & f_{z}(k) P_{galaxy,real-space}(k)\ ,
\end{eqnarray}
where $f_{b}(k)$ is some function describing the bias of the galaxy power
spectrum relative to that of the mass, and $f_{z}(k)$ describes the effect
of redshift distortions caused by galaxy peculiar velocities. In general, the
particular functional forms of $f_b$ and $f_z$ will depend on the parameters 
of the cosmological model. We make the 
simple assumption that over the scales we consider, $f_{b}(k) = b^2 = $
constant, where $b$ is the bias factor. For $f_z(k)$, there are two main 
regimes.
As derived by Kaiser (1987), on large scales in the linear regime, the 
redshift-space power spectrum is amplified by the infall of galaxies from low- 
to high-density regions, and 
\begin{equation} \label{eqbias}
f_z(k) \approx 1 + \frac{2}{3} \frac{\Omega_0^{4/7}}{b}
+ \frac{1}{5} \frac{\Omega_0^{8/7}}{b^2}\ .
\end{equation}
(More commonly in the literature the fractions $4/7$ and $8/7$ in the above
equation are replaced by 0.6 and 1.2 respectively; but see 
\cite{lig90}.)
On the other hand, on small scales in the nonlinear regime, the redshift-space
power spectrum is suppressed by the smearing of the clustering pattern from
small-scale galaxy peculiar velocities. For example, if we assume that galaxies
have a small-scale Gaussian peculiar velocity field with a one-dimensional
dispersion $\sigma_v$, then
\begin{equation}
f_z(k) = \frac{\sqrt{\pi}}{2} \frac{{\rm erf}(k \sigma_v / H_0)}
                                   {k \sigma_v / H_0}
\end{equation}
(\cite{pea92}). However, here we will focus on the linear regime 
and so assume that 
\begin{equation} \label{eqlineartoz}
P_{galaxy,redshift-space}(k) \approx b^2 
\left[ 
1 + \frac{2}{3} \frac{\Omega_0^{4/7}}{b}
+ \frac{1}{5} \frac{\Omega_0^{8/7}}{b^2}
\right] P_{mass,real-space,linear}(k)
\end{equation}
holds as a useful approximation, that is, the redshift-space galaxy power
spectrum is just a constant times the real-space mass power spectrum.

Since the density fluctuations in the galaxy distribution should be less
than unity in the linear regime, and since we observe that the galaxy
fluctuations in $8\ h^{-1}$ Mpc radius spheres $\sigma_8 = 1$, we need
to restrict our fits at least to $k < 2 \pi / ({\rm several} \times 10
\ h^{-1}$~Mpc). We will use $k \leq 0.16 \ h$~Mpc$^{-1}$, corresponding
to $\lambda \geq 40 \ h^{-1}$~Mpc. Though this choice is somewhat
arbitrary, it is motivated by our desire to extend the fit range
beyond the flat part of the convolved power spectrum (to improve the
handle on the power spectrum shape), but not so far that nonlinear
effects become important. That (\ref{eqlineartoz}) approximately
holds over our $k$ range has justification. Theoretical work (e.g.,
\cite{kau95}; \cite{wei95}) indicates that constant linear biasing
does hold on large scales. There is also empirical evidence, from the
comparison of real- and redshift-space galaxy power spectra from the 
survey compilation of Peacock \& Dodds (1994), and in particular, from
the comparison of various galaxy survey power spectra ({\em including} LCRS)
against the {\em mass} power spectrum derived from peculiar velocity
catalogs over the range $0.05 \leq k \leq 0.2 \ h$~Mpc$^{-1}$ (\cite{kol95}).
We will thus fit various model linear power spectra (to be described below) to
the full sample magnitude-limited LCRS {\em convolved} power spectrum
for $0.016 \leq k \leq 0.16 \ h$~Mpc$^{-1}$.
One complication is that because of correlations among the data points
on large scales, a rigorous fit requires us to calculate and account
for the covariance matrix of the errors (e.g., \cite{fel94}). However,
for simplicity we take a more empirical approach and estimate the
errors for the fit parameters from the variation among our usual four
LCRS subsamples and from the N-body mock surveys.

We focus on the case of inflationary, dark-matter dominated
cosmological models, in which case we may write
\begin{equation}
P_{mass,real-space,linear}(k) = B_{mass} k^n T^2(k) \ ,
\end{equation}
where $B_{mass}$ is a normalization factor, $k^n$ is the 
inflation-produced primordial power spectrum of spectral index $n$, and $T(k)$ is a
transfer function that modifies the primordial spectrum. Note that the
shape of the power spectrum is parameterized by the index $n$ and by the
form of $T(k)$, which in turn depends on the specific matter content of
the universe. Now, because we are fitting over a relatively
small range in $k$, and because the convolved power spectrum tends to 
flatten out over the same $k$ range, regardless of the
true power spectrum shape, the LCRS results will not constrain the true
shape very strongly. On the other hand, given a
specific power spectrum shape, the LCRS results do give fairly stringent 
constraints on the power spectrum normalization. Using our earlier
mock surveys and the usual procedure of dividing into four subsamples,
we estimate a $1\sigma$ fractional error of about 10\% in the
normalization, obtained by fitting to the full-sample LCRS power
spectrum over our desired $k$ range. However, there is still freedom
in the choice of bias parameter $b$ (equation~[\ref{eqbias}]) so that
we cannot fix an absolute normalization from our galaxy data. The
absolute mass normalization $B_{mass}$ can instead be determined independently
using results from the COBE DMR experiment.
We can then turn the problem around and use our derived galaxy normalization
$B_{LCRS}$ to determine the value of $b$ needed to reconcile the LCRS
and COBE normalizations via the relation
\begin{equation} \label{eqcobetoz}
b^2 
\left[ 
1 + \frac{2}{3} \frac{\Omega_0^{4/7}}{b}
+ \frac{1}{5} \frac{\Omega_0^{8/7}}{b^2}
\right] B_{mass} \approx B_{LCRS}\ .
\end{equation}
We note that theoretical models suggest a bias $b$ near unity or larger
(e.g., \cite{kau95}). Empirically, we see within the LCRS that 
emission/non-emission samples have bias values within $\approx \pm
15\%$ that of the full
sample (power $\propto b^2$), and that super-$M^*$ galaxies have bias
$\approx 1.3$ times that of fainter galaxies. Likewise,
Peacock \& Dodds (1994) find a relative bias between optical and
IRAS galaxies of $b_{optical}:b_{IRAS} = 1.3$, and Kolatt \& Dekel (1995)
find a range $0.77-1.21$ in relative bias parameters for various
surveys, including LCRS. In particular, the latter authors find
$b_{LCRS} = \Omega_0^{0.6} / (0.99 \pm 0.13)$ for $k = 0.05 - 0.2 \
h$~Mpc$^{-1}$. For $\Omega_0 = 1$, this corroborates the earlier indication from our
luminosity bias results that LCRS galaxies may have $b \approx 1$. In any
event, it appears that $0.7 - 1.3$ may represent a reasonable range
in (absolute) bias values for galaxies, and the assumption that LCRS
galaxies are roughly unbiased should serve as a useful guide to help
constrain the models.

To narrow down the cosmological parameter space to be examined, we
will consider four simple variants of the basic flat, inflationary
CDM model with $\Omega_0 = 1$ and $n = 1$. These are: (1)
$\Lambda$CDM, flat CDM models with non-zero cosmological constant
$\Omega_\Lambda = 1 - \Omega_0$; (2) OCDM, open CDM models with
$\Omega_0 < 1$; (3) CHDM, flat models with both CDM and hot dark
matter in the form of massive neutrinos; and (4) TCDM, flat CDM models
with a tilted primordial spectrum of index $n \neq 1$. These models
were motivated by efforts to address various deficiencies in matching the
``standard'' $\Omega_0 = 1$, $h = 0.5$ CDM model to observational
constraints (e.g., see \cite{lid96a},b,c and references therein).
For all these models we need the basic CDM transfer function 
(\cite{bar86}) 
\begin{equation}
T_{CDM}(q) = \frac{\ln(1+2.34q)}{2.34q} 
             [1 + 3.89q + (16.1q)^2 + (5.46q)^3 + (6.71q)^4]^{-1/4} \ ,
\end{equation}
where $q = k / h \Gamma$ and $\Gamma$ is the so-called shape
parameter. For no baryonic matter $\Gamma$ is simply $\Omega_0 h$,
but more generally (e.g. \cite{lid96c})
\begin{equation} \label{eqgamma}
\Gamma = \Omega_0 h \exp(-\Omega_B - \Omega_B / \Omega_0) \ ,
\end{equation}
where $\Omega_B$ is the fraction of critical density in baryons. We
will adopt the big-bang nucleosynthesis value 
$\Omega_B = 0.015 / h^2$ (\cite{cop95}). As a parenthetical remark,
we note that in previous LCRS papers (e.g., \cite{lan96}; \cite{lin95}) we used a
slightly different CDM transfer function (\cite{bon84}; \cite{efs92})
which has $\Omega_B = 0.03$ and $\Gamma = \Omega_0 h$ rather than the
current definition (\ref{eqgamma}). The difference in $\Gamma$ is only 6\% for
$\Omega_0 = 1$.
For our four cases we have
\begin{equation}
T(k) = 
\left\{
\begin{array}{ll}
T_{CDM}(q) & {\rm \Lambda CDM, OCDM, TCDM} \\
T_{CDM}(q) D(q, \Omega_\nu) & {\rm CHDM}
\end{array}\ ,
\right.
\end{equation}
where $D(q,\Omega_\nu)$ is a somewhat complicated function describing
the effects of massive neutrinos (it is given in detail in 
\cite{pog95} and \cite{lid96b}). For our power spectrum definition,
the necessary COBE normalizations may be expressed via
\begin{equation}
B_{mass} = (2 \pi^2 c / H_0)^{3+n} \delta_H^2 \ ,
\end{equation}
where the four-year COBE DMR data (\cite{ben96}) give 
(\cite{lid96b},c; \cite{via96}; Bunn \& White, in preparation)
\begin{equation}
10^5 \delta_H = 
\left\{
\begin{array}{ll}
1.94 \ \Omega_0^{-0.785-0.05 \ln \Omega_0} & \\
  \ \times \exp[-0.95(n-1)-0.169(n-1)^2]
   & {\rm \Lambda CDM, TCDM, CHDM (\Omega_0 = 1)} \\
1.95 \ \Omega_0^{-0.35-0.19 \ln \Omega_0} 
   & {\rm OCDM}
\end{array}\ ,
\right.
\end{equation}
where the $1\sigma$ fractional uncertainty in $\delta_H^2$ is 15\%
(quoted from \cite{lid96b}).
Also, to provide an additional handle on our parameter space, we consider
the constraints provided by the observed abundance of galaxy clusters.
The cluster results may be expressed in terms of the (linear theory) value
of $\sigma_{8,mass}$ as follows (\cite{via96}; \cite{lid96b},c): 
\begin{equation}
\sigma_{8,mass} = 
\left\{
\begin{array}{ll}
0.60 \ \Omega_0^{-(0.59 - 0.16 \Omega_0 + 0.06 \Omega_0^2)}
   & {\rm \Lambda CDM, TCDM} \\
0.60 \ \Omega_0^{-(0.36 + 0.31 \Omega_0 - 0.28 \Omega_0^2)}  
   & {\rm OCDM} \\
0.60 + 0.2 \Omega_\nu / 3 & {\rm CHDM}
\end{array}\ ,
\right.
\end{equation}
where the 95\% confidence uncertainties are $+32\%$ and $-24\%$ 
(but enlarged by the factor $\Omega_0^{0.26 \log_{10} \Omega_0}$ for
$\Lambda$CDM and by $\Omega_0^{0.17 \log_{10} \Omega_0}$ for OCDM; see \cite{via96}).

We consider each of our four classes of CDM models in turn.

{\it (1) $\Lambda$CDM.}
Here we find a shape parameter $\Gamma = 0.3 \pm 0.1$ (the error is one sdom among
the four subsamples). We plot in Figure~\ref{figlcdm}(a) the resulting
bias $b$ as a function of $\Omega_0$, where the error bands are approximate
$2\sigma$ values obtained by adding in quadrature the estimated
$2\sigma$ fractional uncertainties in the LCRS (20\%) and COBE (30\%) 
normalizations $B$. Four sets of curves are shown, drawn for 
$\Gamma = 0.3, 0.2$, and for fixed Hubble constants $h = 0.5$ and 0.8. 
(We note that a variety of measures of $h$, using
Cepheids, supernovae, Tully-Fisher distances, surface brightness fluctuations, 
and other methods, indicate that $h \gtrsim 0.5$,
and likely of order $0.6-0.8$; see e.g. \cite{jac92},
\cite{fre94}, \cite{sch94}, \cite{rie95}). Recalling our earlier
discussion, we also draw horizontal lines at $b = 0.7$ and $1.3$ to
roughly indicate the range of reasonable bias values.
In Figure~\ref{figlcdm}(b)
we plot the corresponding COBE-normalized $\sigma_{8,mass}$ values 
($2\sigma$ error bands) and the 95\% confidence limits from the cluster
abundance constraints. We find that the 
preferred values $\Gamma = 0.3$ and $0.5 \lesssim h \lesssim 0.8$
indicate $0.4 \lesssim \Omega_0 \lesssim 0.7$, but that these models
are anti-biased, $b \sim 0.5$, and also violate the cluster abundance
constraints. 
On the other hand, for $\Gamma = 0.2$, we do find
models at $\Omega_0 \approx 0.4-0.5$ and $h \approx 0.5$
which are unbiased and which also meet the cluster constraints.
An $\Omega_0 = 1$, $\Gamma = 0.3$ model is unbiased but requires an
unreasonably low $h \approx 0.3$ and is marginal with respect to 
cluster abundances.

{\it (2) OCDM.} The $\Gamma$ fit is the same as in the $\Lambda$CDM
case. We plot the corresponding $\Gamma = 0.3$ and $h = 0.5, 0.8$
curves in the $b$ and $\sigma_{8,mass}$ vs. $\Omega_0$ planes in
Figure~\ref{figocdm}. The most dramatic difference in the open CDM
case vs. the flat $\Lambda$CDM case arises from the much weaker
variation with $\Omega_0$ of the COBE normalization. The parameter
space opens up a bit and for $0.4 \lesssim \Omega_0 \lesssim 0.6$ there are
models which satisfy $\Gamma = 0.3$ and $0.5 \lesssim h \lesssim 0.8$,
are unbiased, and match the cluster abundances.

{\it (3) CHDM.} In Figure~\ref{figchdm} we plot our results, 
$b$ and $\sigma_{8,mass}$ vs. neutrino density $\Omega_\nu$, for
two cases $h = 0.5, 0.8$. The $h = 0.8$ results are strongly
anti-biased, $b \approx 0.3$, violate the cluster abundances, and
therefore appear ruled out. On the other hand, the $h = 0.5$ models
are mildly anti-biased, $b \approx 0.8$, and satisfy the cluster 
abundances for $\Omega_\nu \gtrsim 0.2$. The LCRS fit further
constrains $\Omega_\nu$ to $0.2 \pm 0.1$ 
(the error is one sdom among the four subsamples)
for the case $h = 0.5$.

{\it (4) TCDM.} Here we plot $b$ and $\sigma_{8,mass}$ vs. the spectral index
$n$, again for the two cases $h = 0.5$ and 0.8, in Figure~\ref{figtcdm}.
For $h = 0.5$ and $n \approx 0.75$ the models are unbiased and also
satisfy cluster abundances. The LCRS fit gives $n = 0.7 \pm 0.2$. 
For $h = 0.8$ we instead need $n \lesssim 0.6$ to meet the 
constraints, but this appears to violate the direct COBE DMR fit to
the spectral index, $n = 1.2 \pm 0.3$ ($1\sigma$; \cite{ben96}).

In summary, all the simple CDM variants we have considered can provide
models that simultaneously satisfy the LCRS large scale power spectrum
shape constraints, have Hubble constant $0.5 \lesssim h \lesssim 0.8$,
give an approximately unbiased LCRS galaxy distribution on large scales,
and meet the cluster abundance constraints. 
For each of our four CDM classes, we plot in Figure~\ref{figcobe} specific
example linear power spectra that meet the above criteria:
\begin{equation}
\left.
\begin{array}{lllll}
{\rm \Lambda CDM:} 
            &  \Omega_0 = 0.5 & \Gamma = 0.2     & h = 0.5 & b = 0.9 \\
{\rm OCDM:} &  \Omega_0 = 0.5 & \Gamma = 0.3     & h = 0.65 & b = 0.9 \\
{\rm CHDM:} &  \Omega_0 = 1  & \Omega_\nu = 0.2 & h = 0.5  & b = 0.8 \\
{\rm TCDM:} &  \Omega_0 = 1  & n = 0.7          & h = 0.5  & b = 1.3
\end{array}\ .
\right.
\end{equation}
The model linear power spectra are shown both unconvolved and
convolved, and compared to the corresponding LCRS results. Note the
similarity in the shapes of the convolved power spectra, which we are
actually fitting. However, not surprisingly, the true unconvolved spectra do
in fact differ more in shape, and they also show more power compared to the
directly deconvolved LCRS power spectrum for $\lambda \gtrsim 200 \
h^{-1}$~Mpc. Also shown are corresponding
COBE $1\sigma$ error boxes, shifted to redshift space using
equation~(\ref{eqlineartoz}) and the appropriate bias value.

The acceptable regions of our CDM model parameter space appear to be the
following. Flat $\Omega_0 = 1$, $h = 0.5$ models are still
viable, if we add massive neutrinos with $\Omega_\nu \approx 0.2-0.3$
or a spectral tilt $n \approx 0.7-0.8$. Alternatively, flat $\Omega_0
+ \Omega_\Lambda = 1$ CDM models with $\Gamma \approx 0.2$ and
$\Omega_0 \approx 0.4 - 0.5$ will also do, and similarly for open CDM
models with $\Omega_0 \approx 0.4 - 0.6$ and $\Gamma \approx 0.3$.
We also note that these models can provide $\sigma_{8,mass}$
values near unity, which was suggested in the earlier comparison of
our luminosity bias results with the predictions of Kauffmann et al.\
(1995). Similarly, these models can also meet the constraint
$\sigma_{8,mass} \Omega_0^{0.6} \simeq 0.7-0.8$ from peculiar velocity
data (\cite{kol95}). Additionally, they also satisfy the 95\%
lower bound on the age of the universe, 12 Gyr, as determined from the ages of
the oldest globular clusters (\cite{chab96}).

We have by no means attempted to
exhaustively compare our results to all the possibilities (e.g., we can
incorporate tilt in the open models, add gravity waves, and so on),
but have simply explored some simple variations on the CDM theme, and
have seen that there is already room for a number of such viable models.
The next generation of {\em very} large redshift surveys, such as
Sloan (\cite{gun95}) and 2DF (\cite{ellis93}) should provide
us with improvements on the large-scale galaxy power spectrum
(\cite{vog95}). In conjunction
with improvements in degree-scale microwave anisotropy experments
(e.g., see reviews by \cite{sco95}; \cite{bon95}), the convergence of
the large scale structure and microwave background observational
fronts will hopefully give us a more definitive answer. 
On the other hand, for smaller scales $\lambda < 100 \ h^{-1}$~Mpc, $P(k)$ is
already quite well constrained in the LCRS. In particular, any viable
model must match these small scale results, as we have seen
earlier in the case of our two N-body models.
The complication here is that galaxies exhibit
type-dependent clustering differences (as is the case for our emission 
samples and for our super-$M^*$ galaxies),
which will make it more difficult to interpret which types of galaxies 
are actually being simulated by the models, in as much as the process of
galaxy formation is still not well understood, and there is freedom 
of choice in how one's ``galaxies'' are defined and represented in the
simulations. 
The connection between the N-body models and the statistics of a galaxy
sample will remain a weak link which will require a more
thorough understanding of the types of galaxies selected by
a particular survey and their relation to the underlying
density fluctuations.

\section{Conclusions} \label{ps:conclusions}

We have computed the power spectrum for a magnitude-limited sample of 
19305 Las Campanas survey galaxies, with average redshift $z = 0.1$ and 
absolute magnitudes $-22.5 < M < -18.5$, over scales 
$\lambda = 5 - 400\ h^{-1}$~Mpc. 
$P(k)$ may be approximated by a power law $P(k) \propto k^n$, with slopes
$n = -1.8 \pm 0.1$ on small scales $\lambda = 5 - 30 \ h^{-1}$~Mpc, and 
$n = 1 \pm 1$ on the largest scales $\lambda \approx 200-400\
h^{-1}$~Mpc. The change in slope of $P(k)$ is real and could be the
result of a possible maximum in $P(k)$ observed at $\lambda \approx
100 \ h^{-1}$~Mpc, though our errors preclude a definitive detection
of the turnover.
The overall amplitude may be expressed as $\sigma_8 = 1.0 \pm
0.1$ in redshift space.
We have also compared our results to those from other redshift surveys. For
$\lambda \lesssim  100\ h^{-1}$~Mpc, the LCRS $P(k)$ is similar in
shape to that of the other samples, but with some
differences in amplitude that may reflect intrinsic clustering
differences for galaxies of different luminosities or types (below). For
$\lambda \gtrsim  100\ h^{-1}$~Mpc, the LCRS results are consistent with
those of other surveys given the large errors among all the surveys on
these scales. The overall LCRS results agree best with that from the combined
SSRS2+CfA2 redshift surveys. 

We find evidence in the LCRS for the dependence of clustering strength on galaxy
type. A sample of LCRS emission galaxies with [OII] $\lambda$3727 equivalent 
widths $\geq 5$ \AA \ are about 30\% less clustered than the full LCRS sample
on scales $\lambda \lesssim 100 \ h^{-1}$~Mpc. This may reflect an 
intrinsically weaker clustering strength for spiral galaxies, in qualitative
agreement with the weaker clustering of IRAS vs.\ optically-selected
galaxies.
We also compute the power spectra of volume-limited LCRS samples to 
test for evidence of luminosity bias. We find that LCRS galaxies 
in the range $-18.5 > M - 5 \log h > -21$ appear to be clustered similarly, over scales
$\lambda \approx 5 - 150 \ h^{-1}$~Mpc. However, more luminous galaxies,
which lie about 1 magnitude brighter than $M^*_{LCRS} = -20.3 + 5 \log h$, exhibit
about a 50\% stronger clustering amplitude, in agreement with similar 
results seen in the CfA2 and SSRS2 surveys and the 
qualitative expectations of biased galaxy formation scenarios.

Our full nonlinear
power spectrum is intermediate between that of two CDM N-body models, an 
open $\Omega_0 h = 0.2$ model and a standard $\Omega_0 h = 0.5$ model,
both normalized to $\sigma_8 = 1$ for galaxies. The models agree well
with the LCRS results on small scales $\lambda \lesssim 50 \
h^{-1}$~Mpc, but neither model provides a completely satisfactory fit
to the full LCRS power spetrum. To explore more of model parameter space,
we then fit the large scale LCRS power, on approximately linear 
scales $\lambda \gtrsim 40 \ h^{-1}$~Mpc, to several classes of linear CDM
models. We find that a number of viable models remain which can
simultaneously satisfy constraints provided by the LCRS
power spectrum, the Hubble constant range $0.5 \lesssim h \lesssim
0.8$, cluster abundance results, and the reasonable assumption that
LCRS galaxies are approximately unbiased on large scales, relative to the mass
normalization provided by the 4-year COBE DMR data. These models
include: (1) flat $\Omega_\Lambda + \Omega_0 = 1$ CDM models with
$\Omega_0 \approx 0.4-0.5$ and shape parameter $\Gamma \approx
\Omega_0 h \approx 0.2$; (2) open CDM models with $\Omega_0 \approx
0.4-0.6$ and $\Gamma \approx 0.3$; 
(3) flat $\Omega_0 = 1$ models with CDM plus massive
neutrinos of density $\Omega_\nu \approx 0.2-0.3$; and (4) flat
$\Omega_0 = 1$ CDM models with a spectral tilt $n \approx 0.7-0.8$.

The present situation for power spectrum measurements from redshift surveys
appears to fall into two regimes. On scales $\lambda \lesssim  100\ h^{-1}$~Mpc,
the power spectrum is well determined, and various surveys have consistently
detected differences in the clustering strengths of different galaxy 
populations, though the latter effect may complicate interpretation of model
predictions. On the largest scales $\lambda \gtrsim  100\ h^{-1}$~Mpc, however,
the errors in $P(k)$ are still large. 
We have not been able to conclusively rule out any
of the popular classes of CDM models, though we have begun to constrain
their parameter space. 
Nonetheless, the observational picture should improve
when the next generation of very large redshift surveys come on line, and 
as the degree-scale microwave background observations progress.
In the meanwhile, the Las Campanas results have already provided 
constraints on $P(k)$ on both large and small scales, complementary
to, and largely {\em independent} of, earlier results for smaller and
shallower samples. Furthermore, the type-dependent
clustering differences observed in the survey should provide useful 
observational constraints on models that seek to reproduce accurately the 
clustering properties and formation histories for a realistic mix of 
galaxy types. Toward this end, interested readers are invited to obtain the
Las Campanas Redshift Survey catalog, now publicly available at the
Internet site ``http://manaslu.astro.utoronto.ca/\~{}lin/lcrs.html''.

\acknowledgments

We thank Changbom Park for providing the N-body models, Michael Vogeley
for useful discussions and for providing the SSRS2+CfA2 results, Hume
Feldman for the QDOT results, and Tsafrir Kolatt for helpful comments
on the manuscript.
The Las Campanas Redshift Survey has been supported
by NSF grants AST 87-17207, AST 89-21326, and AST 92-20460. HL also
acknowledges support from NASA grant NGT-51093 and support at the
University of Toronto where this paper was completed.

Note Added in Proof: Power spectrum results for the Stromlo-APM
redshift survey (\cite{tad96}) appeared after submission of the
present paper. Time does not permit a detailed comparison, but
we note that the Stromlo-APM authors report that their results are
consistent with those of the CfA2 101 sample (\cite{par94}) and that
they also find evidence for the stronger clustering of
galaxies brighter than $M^*$.

\clearpage

\begin{deluxetable}{cr}
\tablewidth{30pc}
\tablecaption{Magnitude-Limited Samples\tablenotemark{a}} 
\tablehead{
\colhead{Sample}   & \colhead{$N_{gal}$} 
 }
\startdata
 All                 & 19305 \nl
 North RA $< 12\fh7$ &  5033 \nl
 North RA $> 12\fh7$ &  4327 \nl
 South RA $<  0\fh7$ &  5125 \nl
 South RA $>  0\fh7$ &  4820 \nl
 All 50-fiber        &  4070 \nl
 All 112-fiber       & 15235 \nl
 3727 $W_{\lambda} \geq 5$ \AA & 7712 \nl
 3727 $W_{\lambda} < 5$ \AA & 11593 \nl
\enddata
\tablenotetext{a}{All samples further restricted to 10000 km~s$^{-1}
 < cz <$ 45000 km~s$^{-1}$ and $-22.5 < M - 5 \log h < -18.5$ 
 (``hybrid'' Kron-Cousins $R$ magnitudes).}
\label{ps:tabsamps}
\end{deluxetable}

\begin{deluxetable}{ccr}
\tablewidth{35pc}
\tablecaption{Volume-Limited Samples} 
\tablehead{
\colhead{Absolute Magnitude Limits \tablenotemark{a}} &
\colhead{Distance Limits\tablenotemark{b} ($h^{-1}$ Mpc)} &
\colhead{$N_{gal}$} }
\startdata
 $-18.5 > M > -19.5$ & $ \phn 98 < r < 160$ & 732 \nl
 $-19.0 > M > -20.0$ & $ \phn 98 < r < 198$ & 1867 \nl
 $-19.5 > M > -20.5$ & $ 119 < r < 243$ & 3179 \nl
 $-20.0 > M > -21.0$ & $ 147 < r < 298$ & 4591 \nl
 $-20.5 > M > -21.5$ & $ 181 < r < 362$ & 4404 \nl
 $-21.0 > M > -22.0$ & $ 224 < r < 405$ & 2534 \nl
 $-21.5 > M > -22.5$ & $ 274 < r < 405$ &  582 \nl
\enddata
\tablenotetext{a}{``Hybrid'' Kron-Cousins $R$ magnitudes with $h = 1$.}
\tablenotetext{b}{For nominal 112-fiber apparent magnitude limits 
$15.0 \leq m < 17.7$.}
\label{tabvolsamps}
\end{deluxetable}

\clearpage

\begin{figure} 
\caption{The LCRS galaxy distribution in the Northern and Southern galactic
         caps.}
\label{figslices}
\end{figure} 

\clearpage

\begin{figure}
\plotfiddle{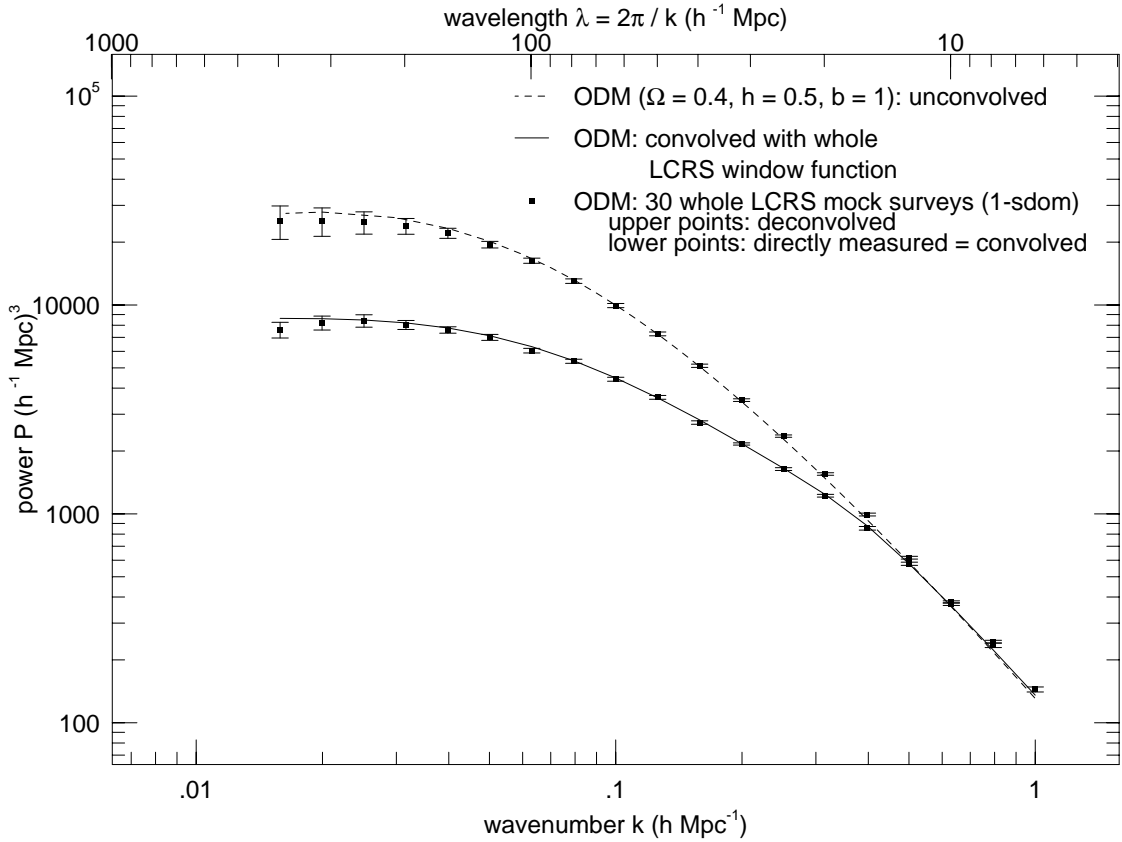}{13cm}{0}{62}{62}{-250}{0}
\caption{The true unconvolved and convolved power spectra of the 
         ODM N-body model,
         and the corresponding power spectra as measured from the ODM 
         LCRS mock surveys. The error bars are one standard deviation of
         the mean (sdom) as measured from the mock surveys.}
\label{figodmsim}
\end{figure} 

\clearpage

\begin{figure} 
\plotfiddle{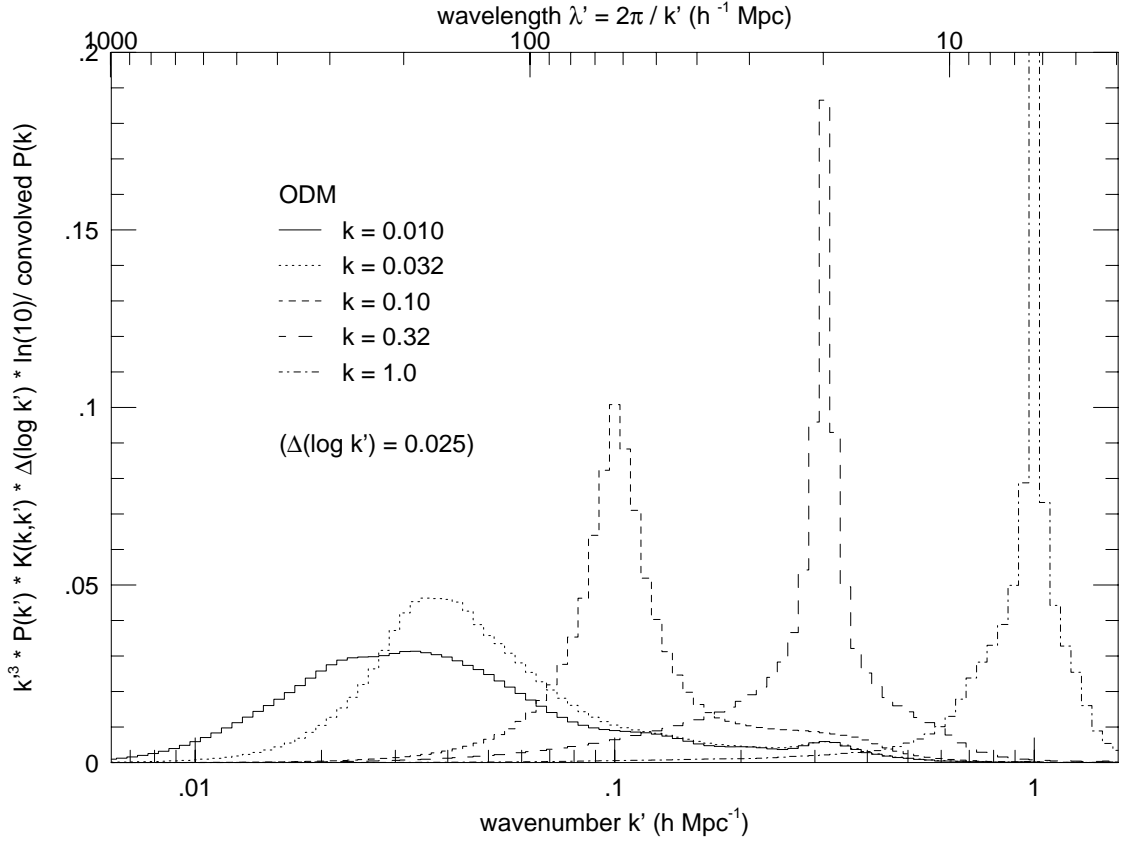}{13cm}{0}{62}{62}{-250}{0}
\caption{Integrand ${k'}^2 P(k') K(k,k')$ of the convolution integral (16) 
         for the ODM model at 5 values of $k$. Note we have multiplied the
         integrand by $k' \ln 10\ \Delta(\log_{10} k')$ as appropriate for our
         log-linear plot, and we have also normalized by dividing by
         $\tilde{P}(k)$. }
\label{figkernel}
\end{figure} 

\clearpage

\begin{figure}
\plotfiddle{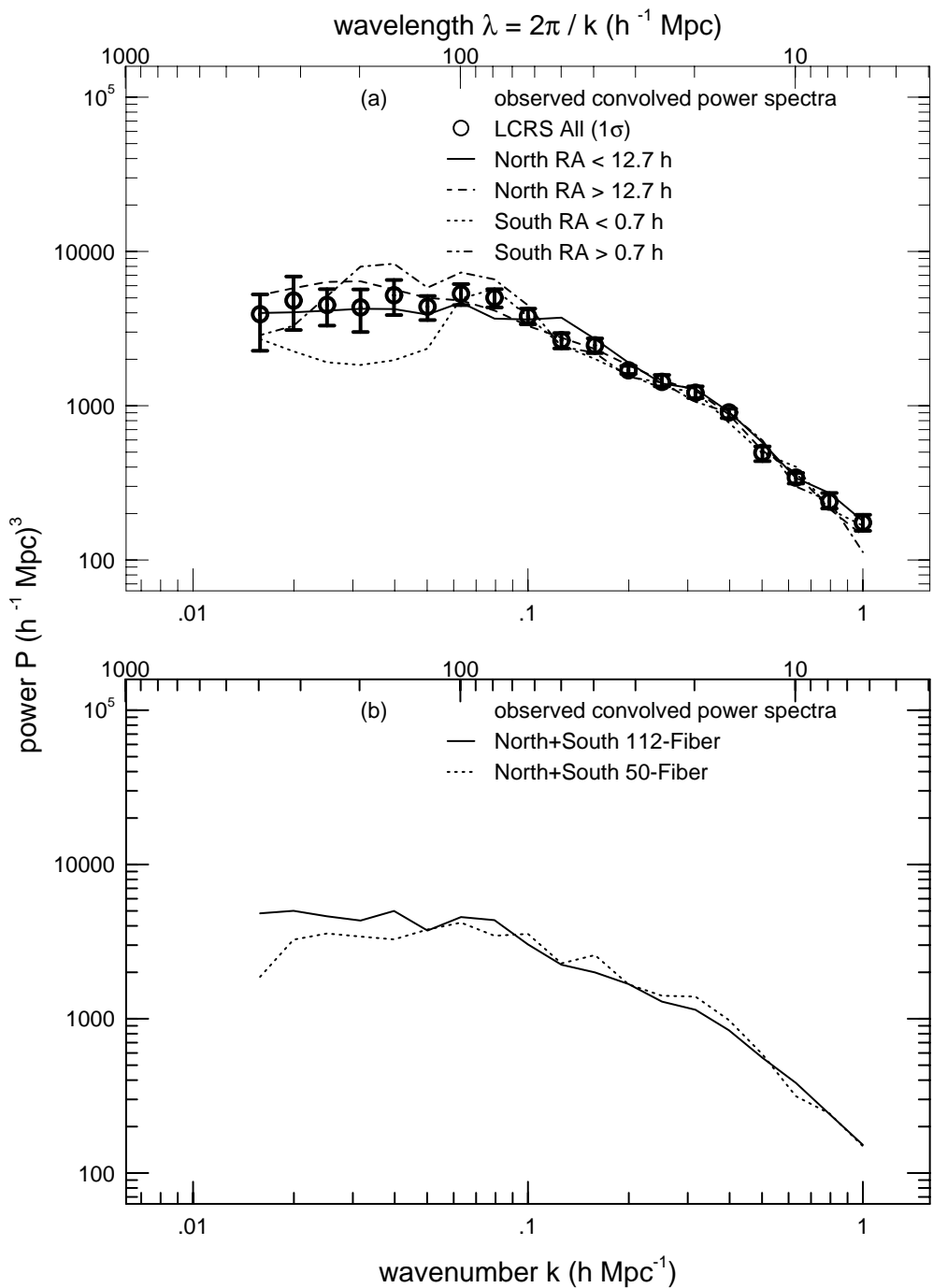}{20cm}{0}{75}{75}{-225}{0}
\caption{(a) The directly measured convolved power spectrum for the 
         full magnitude-limited LCRS
         sample and for each of the four subsets of the full 
         sample divided by hemisphere and right ascension. (b) The convolved
         power spectrum for the 112- and 50-fiber subsets of the full sample.}
\label{fig415}
\end{figure}

\clearpage

\begin{figure} 
\plotfiddle{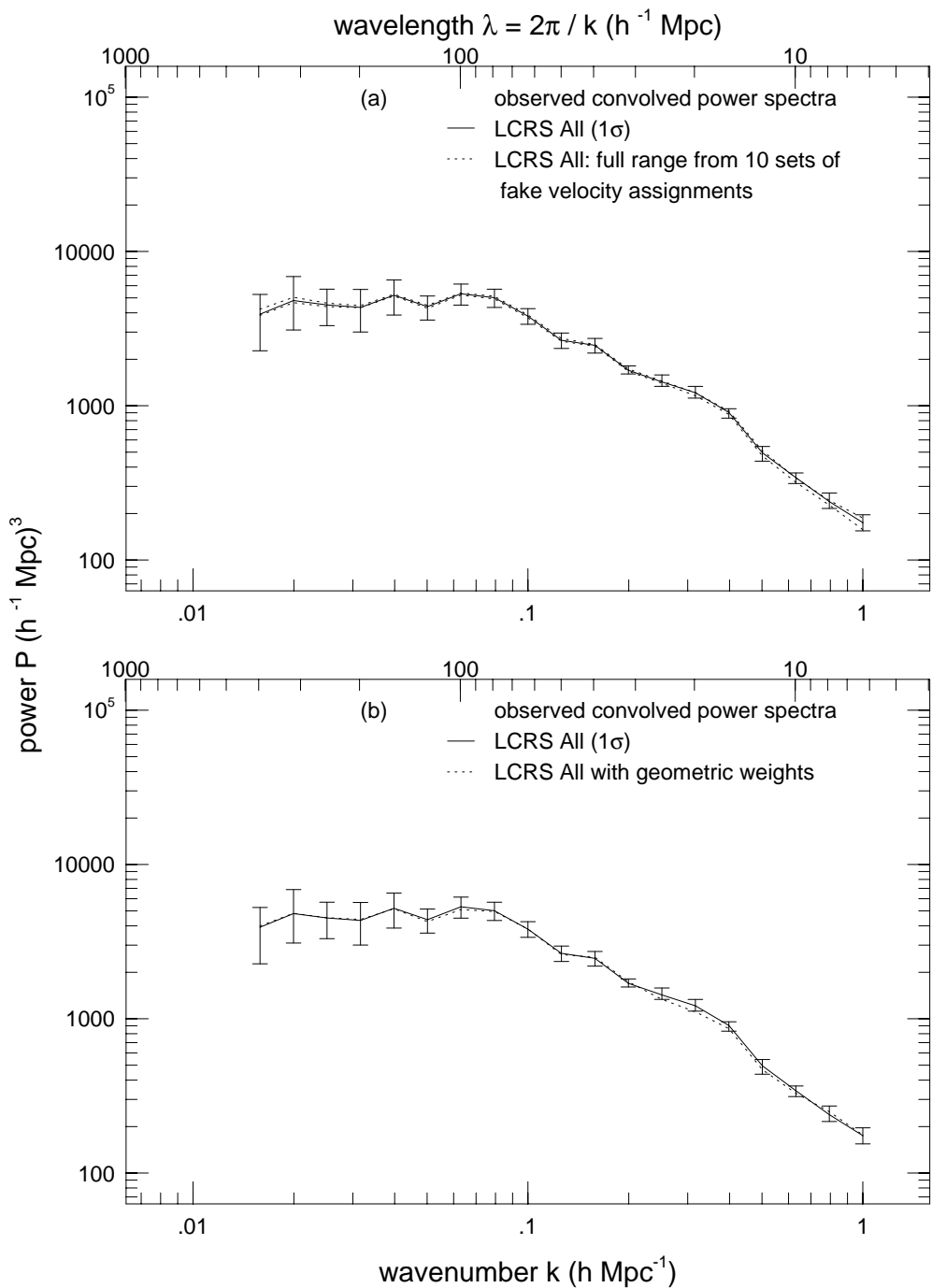}{20cm}{0}{75}{75}{-225}{0}
\caption{The effects on the power spectrum of the full magnitude-limited
         LCRS sample 
         arising from: (a) assigning fake velocities to unobserved
         objects lying within $55\arcsec$ of an observed galaxy;
         and (b) applying geometric weights to correct for the reduced
         spectroscopic success rate at field corners. See text for
         details.}
\label{figgeo}
\end{figure} 

\clearpage

\begin{figure} 
\plotfiddle{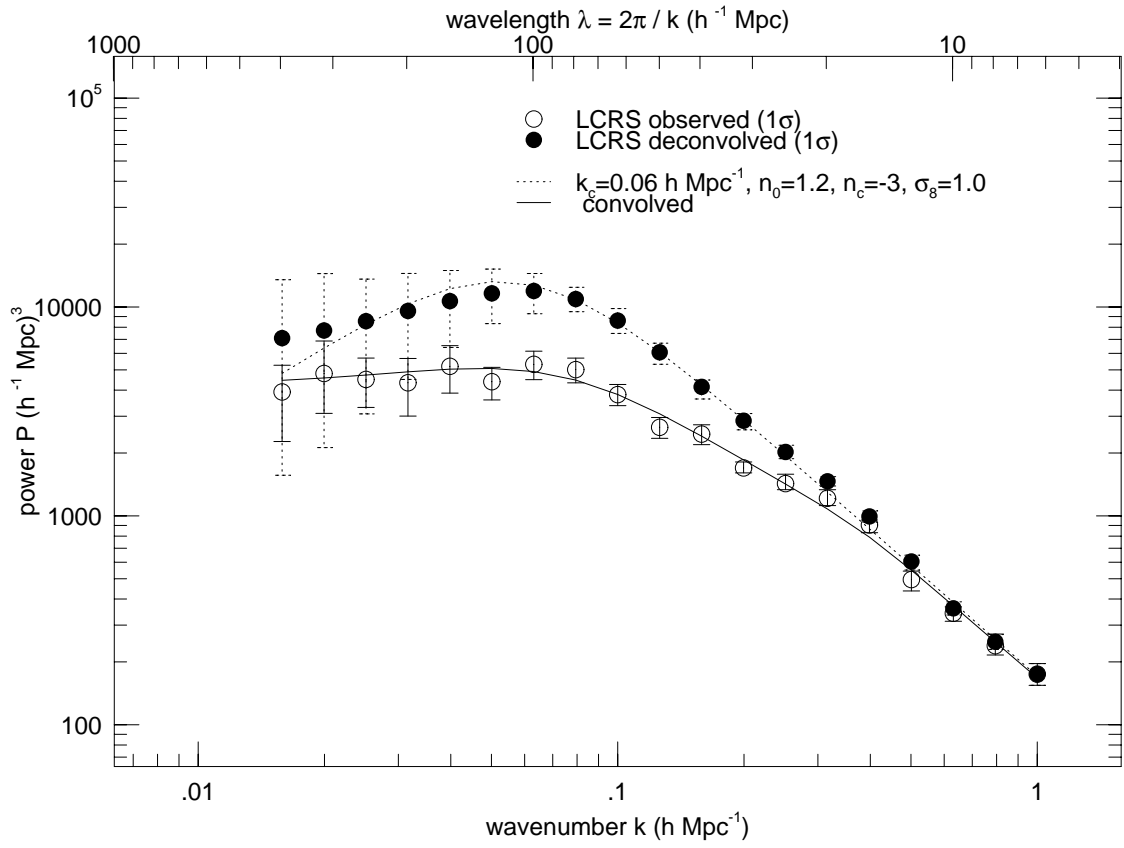}{13cm}{0}{62}{62}{-250}{0}
\caption{The fit to the LCRS power spectrum using the model of
         equation~(23). }
\label{fignsfit}
\end{figure} 

\clearpage

\begin{figure} 
\plotfiddle{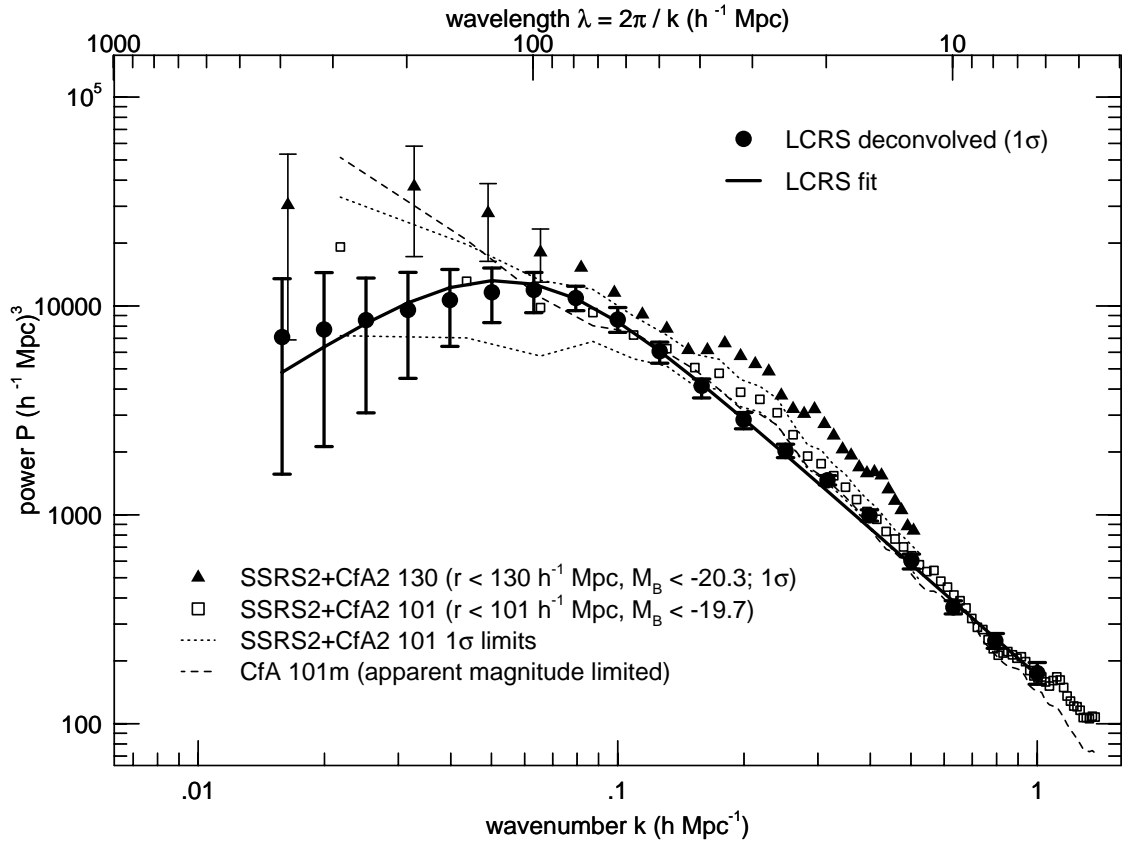}{13cm}{0}{62}{62}{-250}{0}
\caption{The deconvolved LCRS power spectrum and fit compared to the 
         power spectra of three samples drawn from the combined SSRS2 and 
         CfA2 redshift surveys.}
\label{figcfa}
\end{figure} 

\clearpage

\begin{figure} 
\plotfiddle{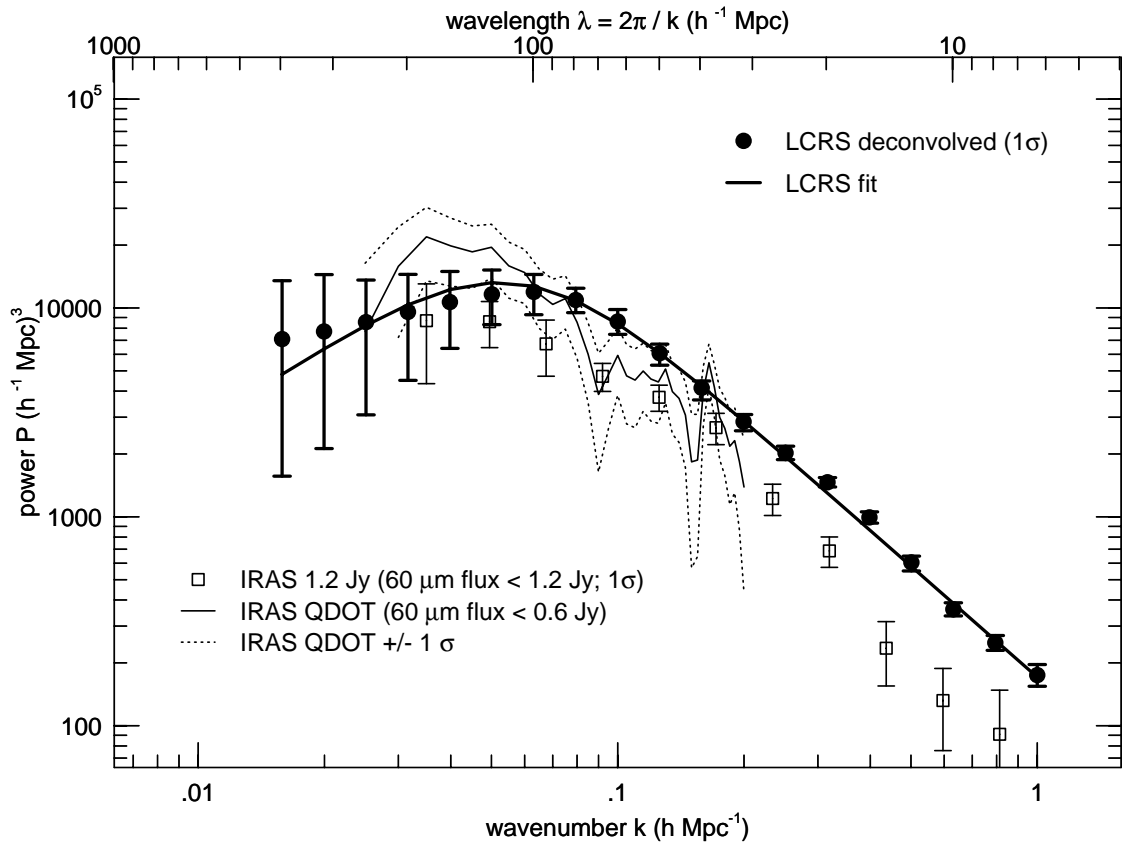}{13cm}{0}{62}{62}{-250}{0}
\caption{The deconvolved LCRS power spectrum and fit
         compared to the power spectra
         of the IRAS 1.2 Jy and IRAS QDOT redshift surveys.}
\label{figiras}
\end{figure} 

\clearpage

\begin{figure} 
\plotfiddle{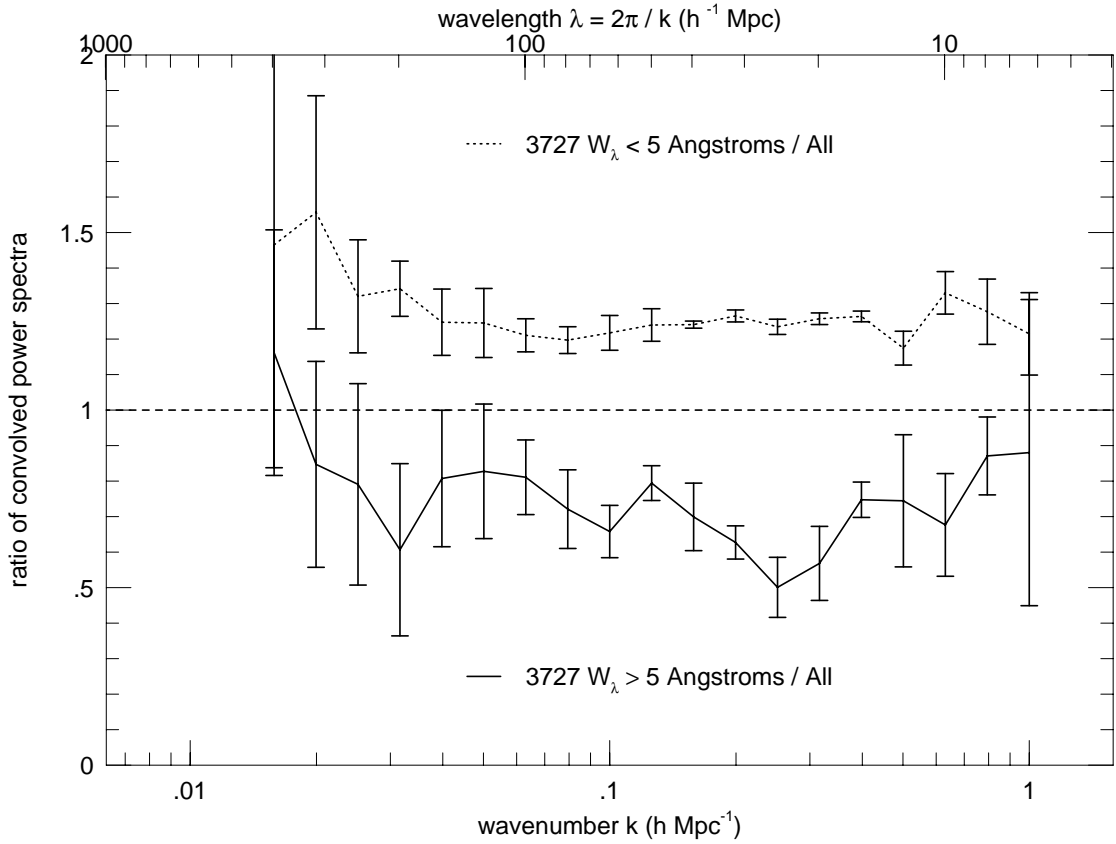}{13cm}{0}{62}{62}{-250}{0}
\caption{The convolved power spectra of 
         two LCRS subsamples, defined by [OII] $\lambda$3727 equivalent width
         $W_{\lambda}$, relative to that of the full sample. 
         The errors shown are standard deviations of the 
         mean determined from the four LCRS subsets divided by hemisphere
         and right ascension.}
\label{figem5}
\end{figure} 

\clearpage

\begin{figure} 
\plotfiddle{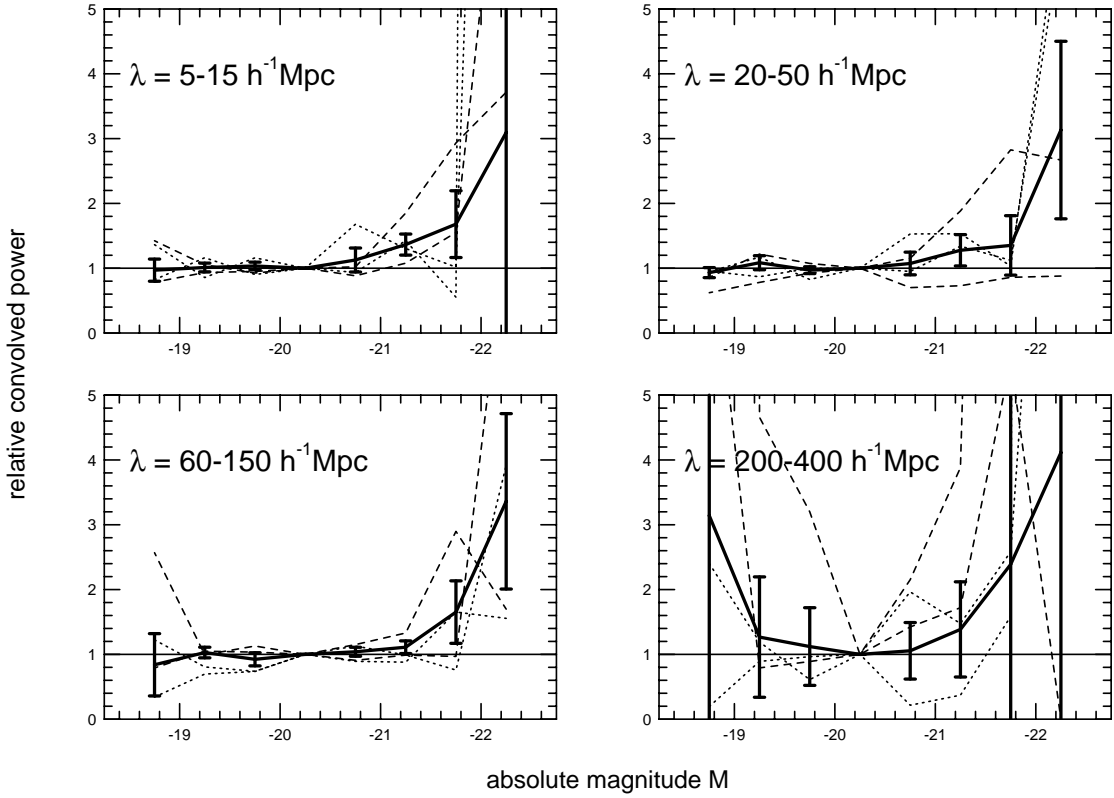}{13cm}{0}{62}{62}{-250}{0}
\caption{The relative convolved power as a function of absolute magnitude, 
         normalized
         to one at $M = -20.25$, shown for the full LCRS sample 
         ({\em heavy solid lines}) and for the four LCRS subsamples divided by
         hemisphere and right ascension ({\em light broken lines}). The errors
         shown are one standard deviation of the mean from the four subsamples.}
\label{figpowrat}
\end{figure} 

\clearpage

\begin{figure} 
\plotfiddle{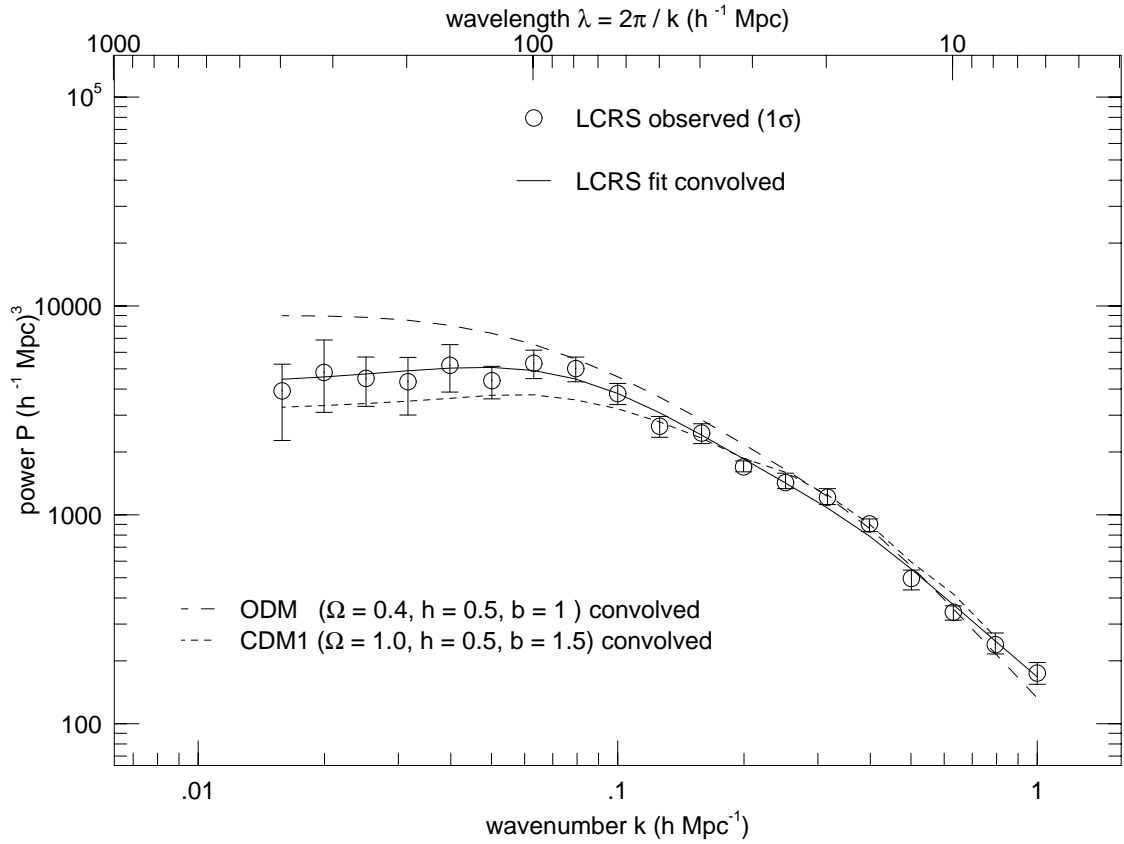}{13cm}{0}{62}{62}{-250}{0}
\caption{The observed convolved LCRS power spectrum compared to the 
         {\em convolved} power spectra of 
         the N-body models ODM and CDM1. See text for details.}
\label{figpnsdm}
\end{figure} 

\clearpage

\begin{figure} 
\plotfiddle{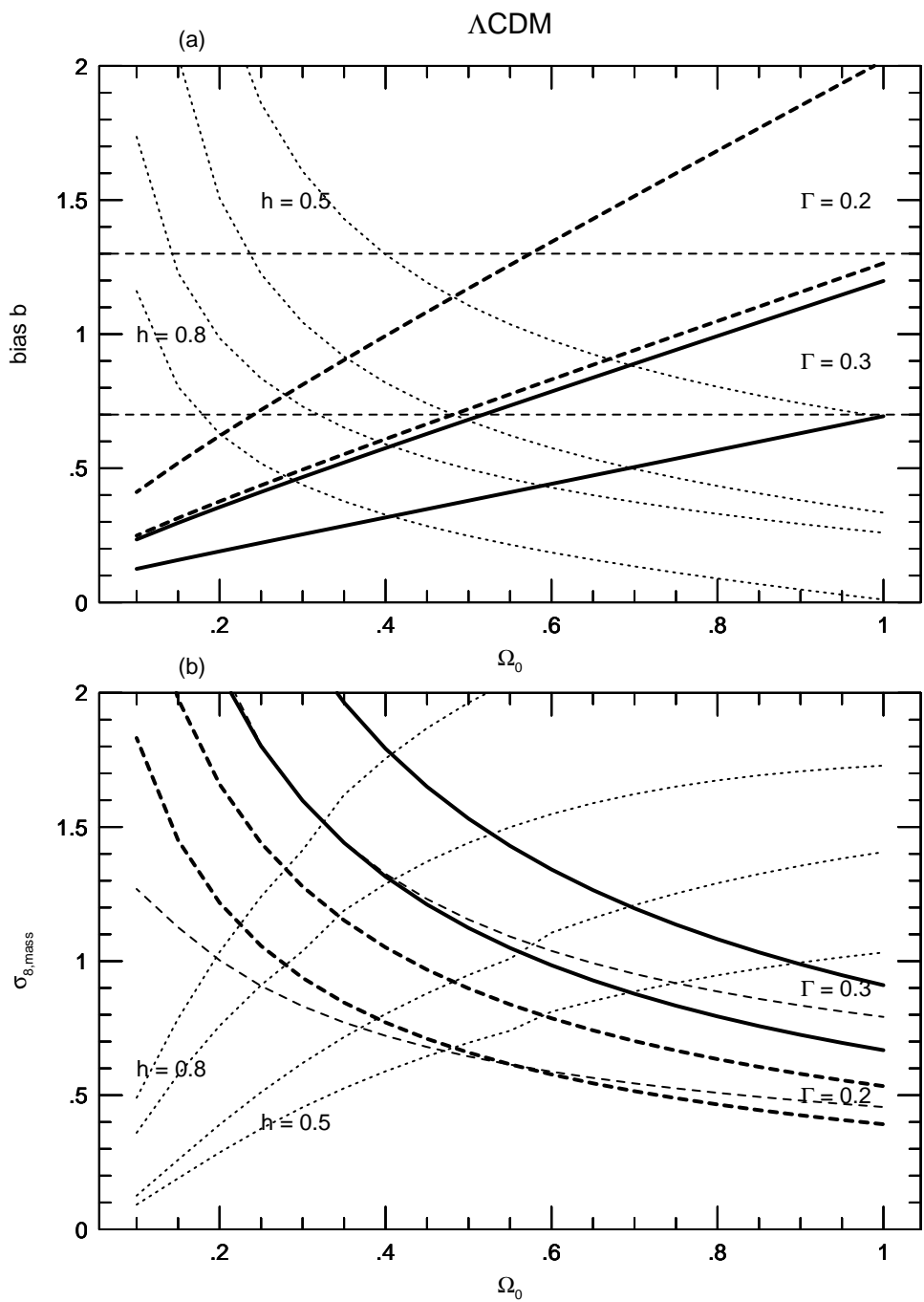}{20cm}{0}{75}{75}{-225}{0}
\caption{(a) Bias $b$ vs. $\Omega_0$ for $\Lambda$CDM models where
             $\Omega_0 + \Omega_\Lambda = 1$. $2\sigma$ error bands
             are shown for fixed shape parameters $\Gamma$ and fixed
             Hubble constants $h$. The lines $b = 0.7$ and 1.3
             roughly indicate the range of reasonable bias values. (b)
             COBE-normalized $\sigma_{8,mass}$ values vs. $\Omega_0$
             for the same models as in (a). The unlabeled thin 
             dashed lines indicate
             95\% confidence limits from cluster abundance
             constraints. See text for more details.}
\label{figlcdm}
\end{figure} 

\clearpage

\begin{figure} 
\plotfiddle{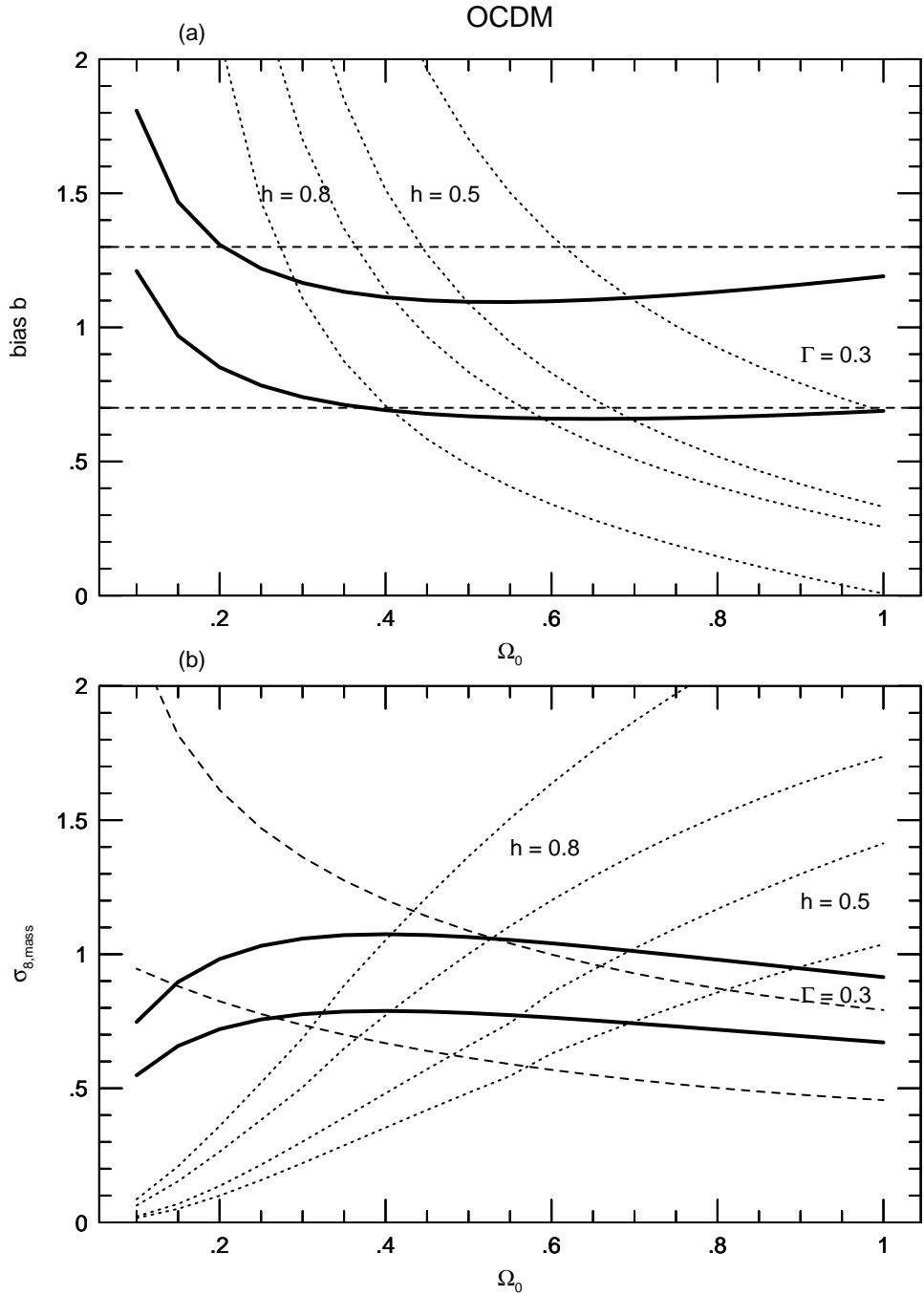}{20cm}{0}{75}{75}{-225}{0}
\caption{Similar to Figure~\ref{figlcdm} except that open CDM models
         with $\Omega_0 < 1$ are being considered.}
\label{figocdm}
\end{figure} 

\clearpage

\begin{figure} 
\plotfiddle{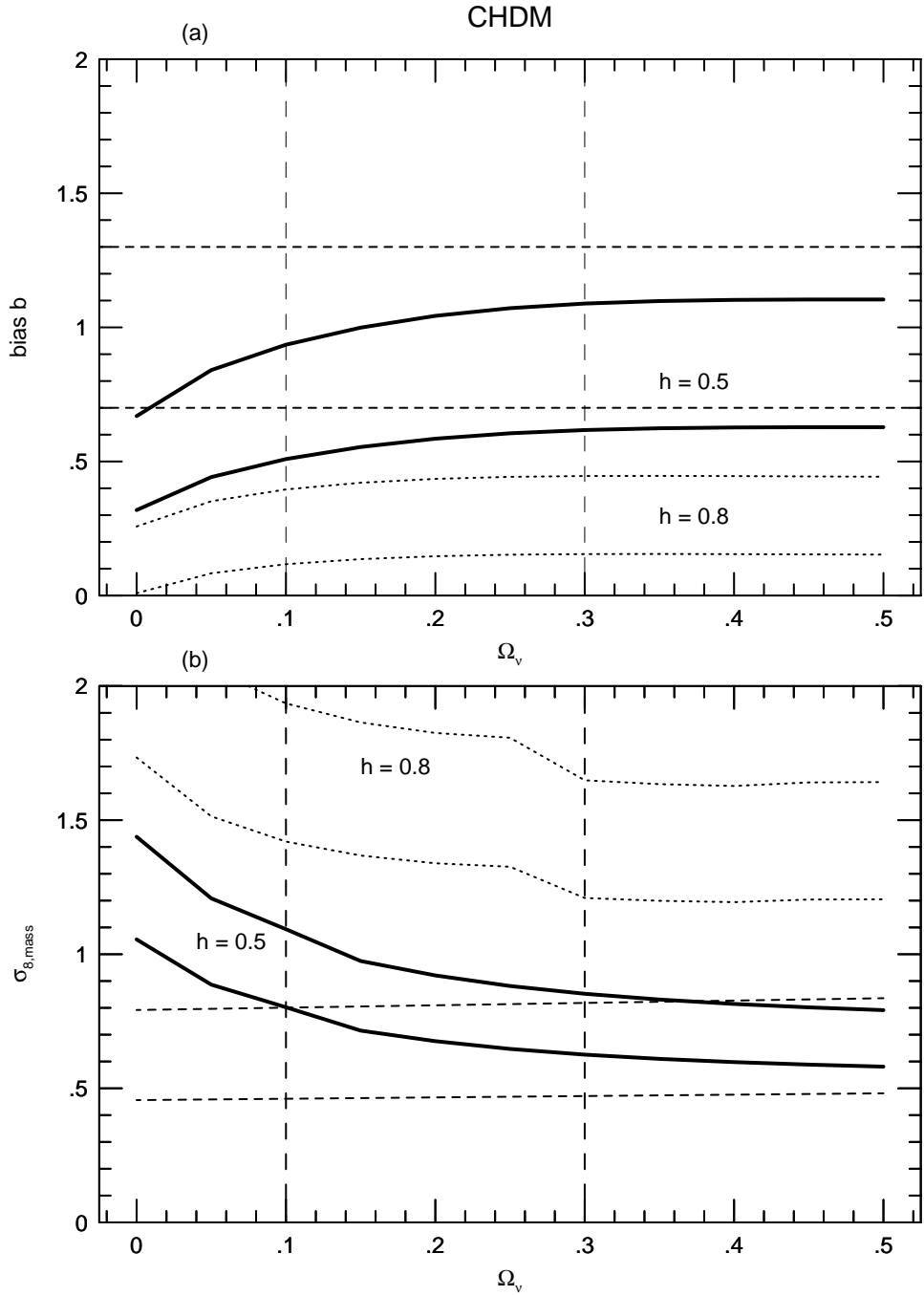}{20cm}{0}{75}{75}{-225}{0}
\caption{Similar to Figure~\ref{figlcdm} except that $\Omega_0 = 1$ CHDM models
         with massive neutrinos of density $\Omega_\nu$ are being
         considered. The vertical lines indicate the LCRS best fit range
         $\Omega_\nu = 0.2 \pm 0.1$ ($1\sigma$) for the case $h = 0.5$.}
\label{figchdm}
\end{figure} 

\clearpage

\begin{figure} 
\plotfiddle{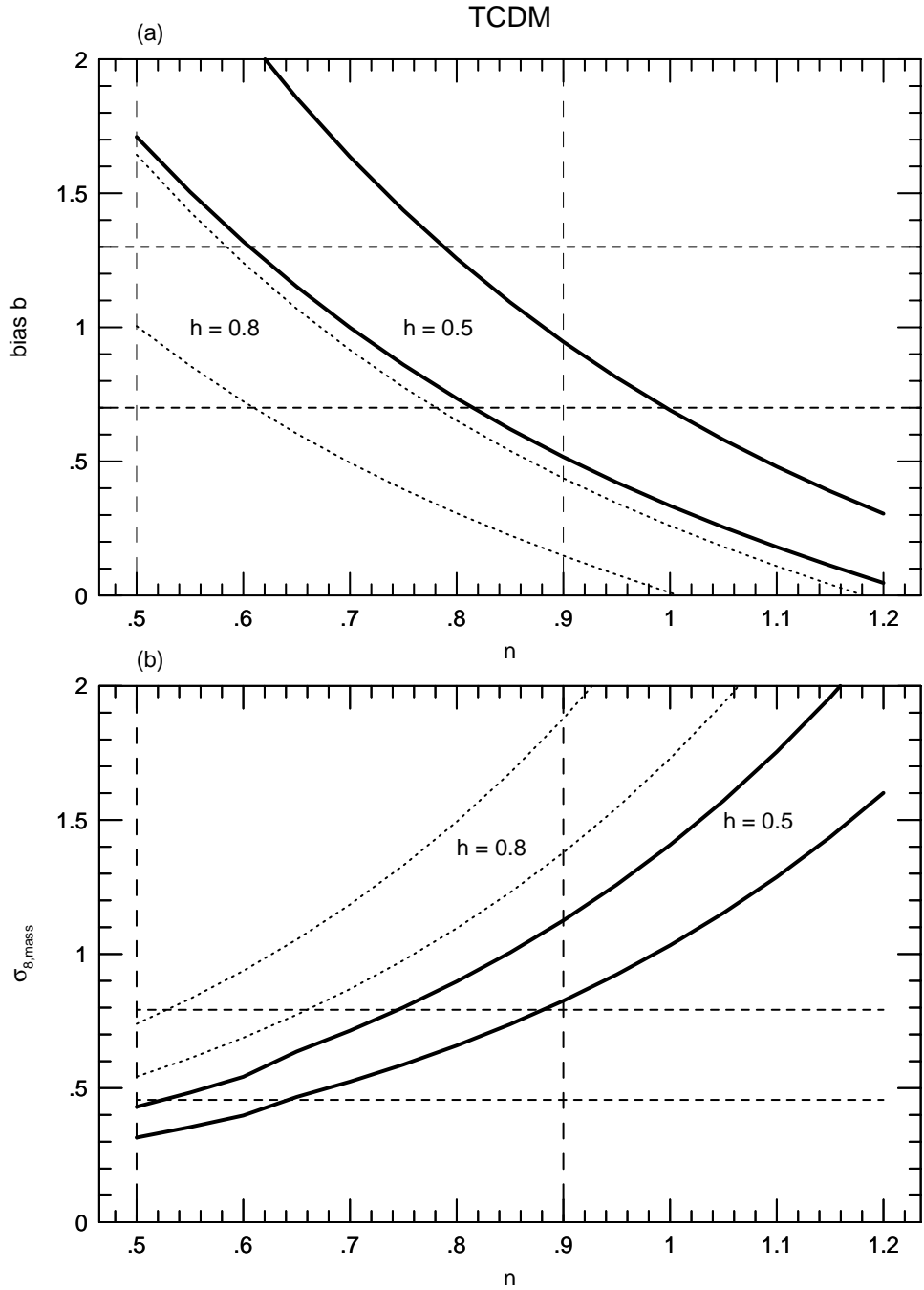}{20cm}{0}{75}{75}{-225}{0}
\caption{Similar to Figure~\ref{figlcdm} except that $\Omega_0 = 1$ tilted
         CDM models with spectral index $n$ are being
         considered. The vertical lines indicate the LCRS best fit range
         $n = 0.7 \pm 0.2$ ($1\sigma$) for the case $h = 0.5$.}
\label{figtcdm}
\end{figure} 

\clearpage

\begin{figure} 
\plotfiddle{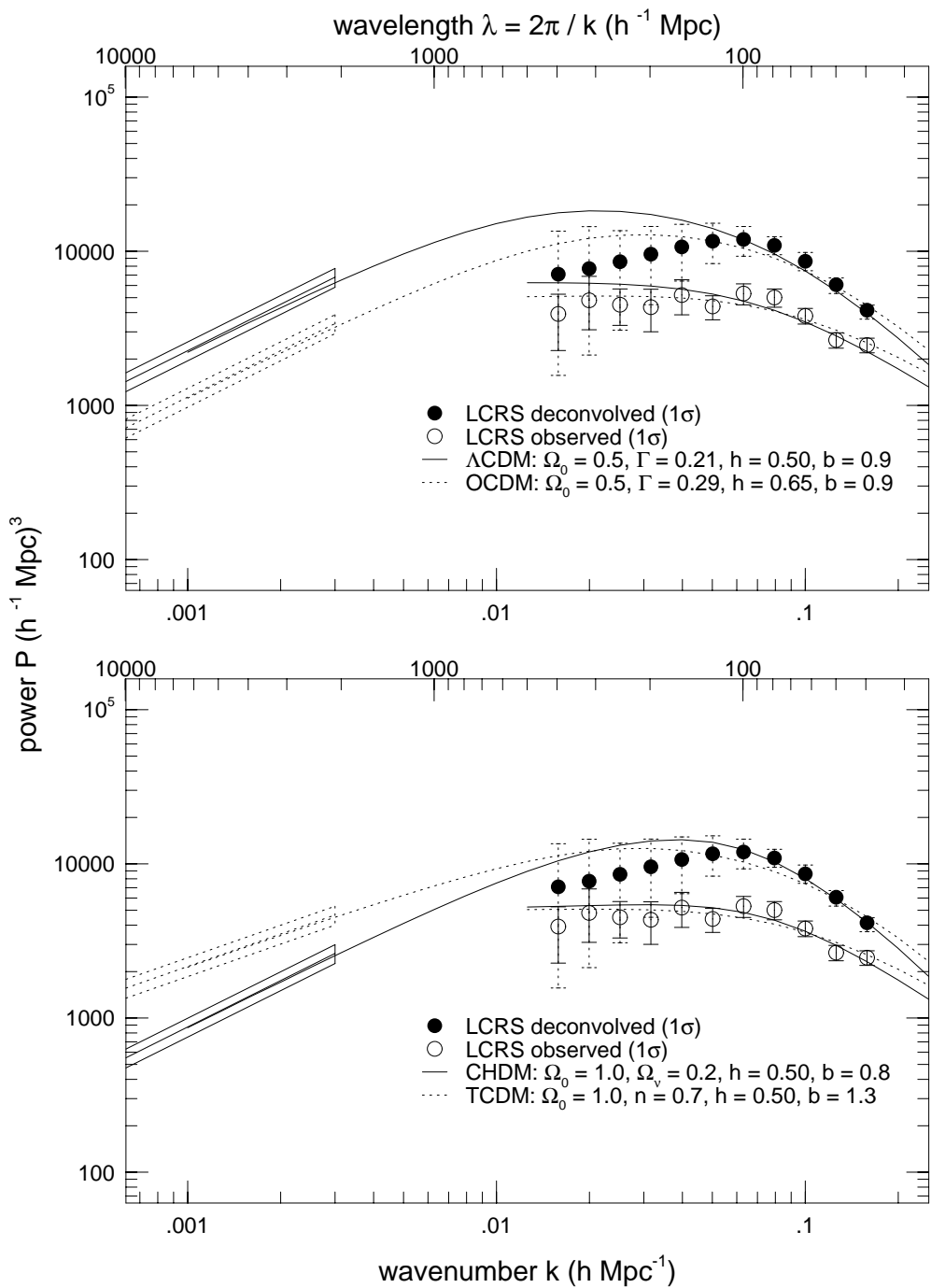}{20cm}{0}{75}{75}{-225}{0}
\caption{The LCRS power spectrum and the fits to it using the linear
         CDM models described in the text, shown both in convolved and
         unconvolved form. Also shown are corresponding COBE $1\sigma$ error boxes,
         transformed into redshift space using the indicated bias
         values. See text for details.}
\label{figcobe}
\end{figure} 

\clearpage

\end{document}